\documentclass[%
 reprint,
superscriptaddress,
 amsmath,amssymb,
 aps,
prx,
longbibliography
]{revtex4-2}

\usepackage{graphicx}
\usepackage{dcolumn}
\usepackage{bm}
\usepackage[hyperfootnotes=false]{hyperref}
\hypersetup{
    colorlinks=true,
    linkcolor=blue,
    citecolor=blue,
    filecolor=magenta,      
    urlcolor=blue,
    pdffitwindow=true
    }
\usepackage[all]{hypcap} 

\makeatletter
\def\@email#1#2{%
 \endgroup
 \@AF@join{#1\texttt{#2}}}
\makeatother

\newcommand{\Leiden}{\affiliation{Laboratory of Experimental Cardiology, Department of Cardiology, Heart Lung Center Leiden, Leiden University Medical Center, Leiden, the Netherlands}}
\newcommand{\Gent}{\affiliation{Department of Physics and Astronomy, Ghent University, Ghent, Belgium}}
\newcommand{\Ekaterinburg}{\affiliation{Ural Federal University, Ekaterinburg, Russia}}
\newcommand{\Sechenov}{\affiliation{World-Class Research Center "Digital biodesign and personalized healthcare", Sechenov University, Moscow, Russia}}
\newcommand{\MaxPlanck}{\affiliation{Laboratory for
Fluid Physics, Pattern Formation
and Biocomplexity, Max Planck
Institute for Dynamics and Self-Organisation, G\"ottingen}}
\newcommand{\Edinburgh}{\affiliation{Institute for Adaptive
and Neural Computation,
Informatics Forum, School of
Informatics, the University of
Edinburgh, Edinburgh, United
Kingdom}}
\begin{document}

\title{Unconventional Collective Resonance as Nonlinear Mechanism of Ectopic Activity in Excitable Media}


\author{Alexander S. Teplenin}
\Leiden

\author{Nina N. Kudryashova}
\Edinburgh

\author{Rupamanjari Majumder}
\Leiden
\MaxPlanck

\author{Antoine A. F. de Vries}
\Leiden

\author{Alexander V. Panfilov}
\email{Alexander.Panfilov@uGent.be$^{1,4,5,6}$}
\Leiden
\Gent 
\Ekaterinburg
\Sechenov

\author{Dani\"el A. Pijnappels}
\email{D.A.Pijnappels@lumc.nl$^{1}$}
\Leiden

\begin{abstract}

Many physical, chemical and biological processes rely on intrinsic oscillations to employ resonance responses to external stimuli of certain frequency. Such resonance phenomena in biological systems are typically explained by one of two mechanisms: either a classical linear resonance of harmonic oscillator, or entrainment and phase locking of nonlinear limit cycle oscillators subjected to periodic forcing. Here, we discover a fundamentally different nonlinear mechanism in a generic family of spatially-distributed systems, which does not require  intrinsic oscillations.
Instead, the resonant frequency dependence arises from coupling between an excitable and a monostable region of the medium. This composite system is endowed with emergent bistability between a stable steady state and stable spatiotemporal oscillations. The resonant transition from stable state to oscillatory state is induced by  waves of particular frequency travelling through the medium.  This transition to the spatiotemporal oscillatory state requires accumulation of multiple waves, resulting in the exclusion of lower frequencies. The cutting off of high frequencies is realized by damping of wave amplitude in the monostable zone and then by activating amplitude sensitive dynamics in the monostable units.
Together, these processes produce a resonance phenomenon that is fundamentally collective in nature. This finding is in stark contrast to what happens in the majority of biophysical systems, where resonant properties are always present at the level of elementary units. 
We demonstrate that this new resonance mechanism can be captured in a simplistic reaction-diffusion model. Also, we reveal this collective resonance mechanism in in-vitro experiments and detailed biophysical simulations representing a major type of arrhythmia, namely ectopic activity in pathological cardiac tissue. 
We further demonstrate, both experimentally and theoretically, that the ongoing spatiotemporal oscillations, such as ectopic activity in cardiac tissue, can be stopped by travelling waves of high frequency. These waves force the excitable medium outside the resonance window, thereby realizing a reverse transition to a stable steady state. This observation provides new theoretical ground for the efficacy of high frequency pacing for termination of cardiac arrhythmia and, due to generality of the proposed termination mechanism, could be utilized more broadly for the control of synchronized oscillatory activity in nonlinear systems.
 Overall, we claim the universality of the presented resonance mechanism in a broad class of nonlinear biophysical systems. Specifically, we hypothesize that such phenomena could be found in neuronal systems as an alternative to traditional resonant processes. Altogether, we have discovered a novel, fundamentally collective resonance mechanism in excitable systems. We demonstrate this mechanism theoretically and empirically in cardiac tissue and show its utility for controlling the behaviour of nonlinear spatially-distributed systems.

\end{abstract}


\maketitle

\section{Introduction}
A variety of complex biological systems employ intermediate frequency-dependent behaviour for different purposes: the actomyosin cortex for actin polymerization \cite{westendorf2013actin}, synthetic biological circuits to express specific genes \cite{guantes2006dynamical}, genetic oscillators to synchronize biological clocks of cells \cite{mondragon2011entrainment} and neurons \cite{hutcheon2000resonance,IZHIKEVICH2000} to fire an action potential or induce spiking activity. These resonance-like phenomena are attributed to one of two mechanism: either a classical resonance of harmonic oscillator \cite{samoilov2002signal}, or entrainment and phase locking of limit cycle oscillators \cite{Murugan_2021,mondragon2011entrainment}.
 Both mechanisms rely on the existence of the system's intrinsic oscillations, which interact with the external stimuli producing frequency-dependent effects. Entrainment and phase locking of  limit cycle oscillator need a properly tuned combination of amplitude and frequency of external stimuli \cite{strogatz2018nonlinear,Murugan_2021}. The classical resonance phenomenon is simpler, since it just requires the system's response to the specific frequency band while attenuating higher and lower frequencies. Also, this mechanism does not require development of sustained intrinsic oscillations, damped oscillations are sufficient. The frequency of external stimulation should be just in resonance with the eigenfrequency of  damped intrinsic oscillations of the system in order to initiate a desired response. In complex biological systems damped oscillations are usually created by putting the stable resting state of the system close to the onset of so-called oscillatory(Hopf) instability \cite{Murugan_2021}. The majority of systems employ all these mechanisms of intermediate frequency selection at the level of a single unit (i.e. a single cell) by developing individual sustained oscillations or by creating damped oscillations via bringing the single unit on the verge of oscillatory instability.
In the current paper, we show that such resonance-like property can be realized in a biological system, namely cardiac tissue, via a completely different mechanism. First, this mechanism does not rely on intrinsic oscillations or does not involve proximity of the resting state to oscillatory instability. Second, it is a purely emergent phenomenon that is not reducible to individual cell behaviour. \par
Generally speaking, the heart can be viewed as a frequency-dependent system but with the negative connotation. In healthy conditions, self-oscillatory activity of single cells at the natural pacemaking region of the heart, the sinus node, emits periodic waves to the remainder of the cardiac tissue, thereby determining the rhythm of the whole heart.  However, in pathological situations, periodic waves from the sinus node can induce life-threatening arrhythmias in other regions of the heart and induction of such arrhythmias is frequency-dependent \cite{qu2014nonlinear}.  
In other words, the emergence of arrhythmias can be viewed as a response to the "input" frequency at the sinus node.
Such arrhythmias can be divided into two main types: vortex-driven arrhythmias and arrhythmias sustained by so-called ectopic activity. The latter is the periodic or transient emission of concentric excitation waves from some point in the heart. Interestingly, ectopic activity can lead to induction of vortex-driven arrhythmias \cite{gong2007mechanism}. Thus, uncovering the mechanisms underlying ectopic activity is of great importance. Induction of vortex-driven arrhythmias is a purely collective phenomenon and is observed only at high, but not at intermediate frequencies of cardiac rhythm \cite{qu2014nonlinear}. In other words, vortex-driven arrhythmias are devoid of resonance-like mechanism of induction. On the other hand, ectopic activity can be evoked in a resonance-like manner \cite{Et2009}, i.e. not inducible by high or low frequency stimulation. This known mechanism of frequency-dependence can be explained by properties of dynamics at the level of a single unit (i.e. a single cardiac cell) without the involvement of collective effects. Such single cell dynamics, which is named early afterdepolarizations, arises due to Hopf bifurcation and exhibits chaotic behavior at intermediate frequencies of stimulation \cite{vo2019pacing,Tran2009,Et2009}.

Here, we find a situation in which a transition to ectopic activity can be induced in a resonance-like manner due to purely collective mechanisms. We do so by capitalizing on our previous results, where we show that persistent ectopic activity can also be a result of spatial dynamics of the system, when no self-oscillatory cells are present \cite{teplenin2018paradoxical}. In the current paper we break new ground by revealing the phenomenon of frequency-dependent transition to ectopic activity as a fundamentally new insight in the collective mechanisms of arrhythmia induction.
To prove it experimentally, we use monolayers of heart muscle cells expressing the light-sensitive depolarizing  ion channel CheRiff  (channelrhodopsin family) \cite{hochbaum2014all}. With this optogenetic tool the strength of the ionic current through this channel can be regulated by intensity of light. Using such methodology,  referred to as optogenetics, it is possible to gradually change the balance between inward and outwards current in the cells, which is similar to electrical alterations during arrhythmogenic pathology. We generated this current regionally by changing  the background current by regulating the intensity of the light.  
This allowed us to perform detailed experimental analysis  of the onset of  ectopic activity under periodic external stimulation.
We confirmed our findings \textit{in-silico} allowing in-depth investigation of the underlying mechanisms.  
 In particular, onset of ectopic activity depends not only on the degree of depolarization, but also on the number and frequency of the waves propagating through the tissue demonstrating the phenomena of  bistability,  multistability and resonance.
 \par
 As a consequence of this resonance phenomenon, we found that ongoing ectopic activity can also be eliminated by external stimulation of higher than the resonant frequency. This phenomenon can extend mechanisms explaining success of applying so-called fast anti-tachycardia pacing for termination of slow ventricular arrhythmias in clinical practice. Previously, the successful application of anti-tachycardia pacing was mechanistically justified only for the class of vortex-driven arrhythmias \cite{qu2014nonlinear,agladze_moving_spiral}. Now we demonstrate such mechanism for ectopic arrhythmia. On a more general level, our termination mechanism might be an alternative scenario for ceasing of oscillations in various media by application of external stimuli in comparison to classically known Winfree's pacemaker annihilation mechanism \cite{winfree2001geometry} or the recently proposed mechanism based on antiresonance regime \cite{PhysRevX.10.011073}.
 \par
 In terms of generalizing our findings we were able to reproduce all observed phenomena \textit{in-silico} both in a detailed model and in a simplified three-variable reaction-diffusion system. As a result, we propose a general mechanism explaining the resonance phenomenon based on bifurcation theory.  In particular, the collective resonance phenomenon arises from coupling between an excitable and a monostable regions of the medium. On such level of description, the monostable region corresponds to  optogenetically depolarized area and the excitable region refers to non-depolarized part in \textit{in-vitro} experiments. As mentioned above, this composite system allows bistability, namely bistability between spatiotemporal oscillatory state (ectopic activity) and stable steady state (non-arrhythmic state). In phase space these states are two dynamical attractors, which are separated by the threshold (threshold manifold). The transition to the spatiotemporal oscillatory state requires accumulation of multiple waves to cross the threshold resulting in exclusion of lower frequencies. The cutting off of high frequencies happens due to decrease of wave amplitude in monostable zone and then by activating amplitude sensitive dynamics in monostable units. We further uncover that those monostable units, if subjected to coupling current of their neighbours, possess "hidden" bistability. Introduced by us in this paper, "hidden" bistability is a novel general nonlinear phenomenon. It allows coexistence of an excitable and bistable regime in a single unit. Switching between these regime depends on the amplitude of external forcing that explains previously mentioned amplitude sensitive dynamics in monostable units. This "hidden" bistability phenomenon is independent of the precise details of the system thus making the phenomenon of collective resonance widely applicable. The resulting generality of the resonance effect makes it interesting not only for the field of cardiac arrhythmias, but also for other resonance phenomena in biological, chemical and physical  nonlinear systems.

 \par
 Our paper is structured as follows. First, we outline the methods in the next section \ref{sec:methods}, then show the general experimental setting of the phenomenon in  Sec. \ref{sec:general_setting}  and enumerate possible dynamical states of the system in Sec. \ref{sec:bifurcation}. Next,  resonance phenomena in the induction of ectopic activity is presented in Sec. \ref{sec:decription_resonance}, the termination phenomenon is introduced in Sec. \ref{sec:termination}. Collective nature of the phenomenon is proved directly \textit{in-vitro} and \textit{in-silico} in Sec.  \ref{sec:collective_nature}. The mechanism is explained in Sec. \ref{sec:mechanism}. In particular, section \ref{sec:two_attr} shows that the mechanism does not involve traditional resonance mechanisms by demonstrating mutual placement of states of the system in the phase space. Section \ref{sec:hidden_bist} describes the "hidden" bistability phenomenon and its involvement in the collective mechanism. Section \ref{sec:minimal_RD} shows results from minimalistic reaction-diffusion model, while section \ref{sec:discussion} is dedicated to discussion.
\section{Methods} 
\label{sec:methods}

 \subsubsection{Preparation of  monolayers of ventricular myocytes expressing light-gated ion channels}
 All animal experiments were reviewed and approved by the Animal Experiments Committee of the Leiden University Medical Center and conformed to the Guide for the Care and Use of Laboratory Animals as stated by the US National Institutes of Health. Monolayers of neonatal rat ventricular cardiomyocytes (NRVMs)  were established as previously described \cite{teplenin2018paradoxical}. Briefly, the hearts were excised from anesthetized 2-day-old Wistar rats. The ventricles were cut into small pieces and dissociated in a solution containing $450{~\rm U/ml}$ collagenase type I (Worthington, Lakewood, NJ) and 18,75 Kunitz/ml DNase I (Sigma-Aldrich, St. Louis, MO). The resulting cell suspension was enriched for cardiomyocytes by preplating for 120 minutes in a humidified incubator at $37^\circ$C and $5$\% CO$_{2}$ using Primaria culture dishes (Becton Dickinson, Breda, the Netherlands). Finally, the cells were seeded on round glass coverslips  ($d=15{~\rm mm}$; Gerhard Menzel, Braunschweig, Germany) coated with bovine fibronectin (100 $\mu$g/ml; Sigma-Aldrich) to establish confluent monolayers. After incubation overnight in an atmosphere of humidified $95$\% air- $5$\% CO$_{2}$ at $37^\circ$C, these monolayers were treated with Mitomycin-C (10 $\mu$g/ml; Sigma-Aldrich) for 2 hours to minimize proliferation of the non-cardiomyocytes. At day 4 of culture, the NRVM monolayers were incubated for 20-24 h with lentiviral vector particles encoding CheRiff, an light-gated depolarizing ion channel of the channelrhodopsin family, at a dose resulting in homogeneous transduction of nearly $100$\% of the cells. Next, the cultures were washed once with phosphate-buffer saline, given fresh culture medium and kept under culture conditions for 3-4 additional days.\\

\subsubsection{Detailed electrophysiological model}

In this study, the  Majumder-Korhonen model of NRVM monolayers was \cite{majumder2016islands}, the monodomain reaction-diffusion equation is formulated as follows:
 \begin{eqnarray}
   \frac{\partial V}{\partial t} = \nabla \cdot (\mathcal{D} \nabla V) - \frac{I_{ion}+I_{ChR2}(x,y)}{C_m},
 \end{eqnarray}
where $I_{ion}$ is the sum of total ionic currents, $D$ is the  diffusion coefficient equal to 0.00095 cm$^2$/ms and $I_{ChR2}(x,y)$ is the spatially controlled optogenetic current. In comparison to the  original model, $I_{Na}$ was increased $1.3$-fold and, $I_{Ca_L}$ and $I_{Kr}$ were both reduced $0.7$-fold. The resulting conduction velocity (CV) was 20 cm/s.
The steady-state voltage dependence for the inactivation variable $h$ of the fast Na$^{+}$ current was changed to $h_{\infty}=[ {1+e^{\frac{(72+V)}{6.07}}}]^{-1}$, where $V$ is the transmembrane potential. The optogenetic current $I_{ChR2}$ was formulated based on the Boyle $I_{ChR2}$ model \cite{boyle2013comprehensive} with the introduction of rectifiying conduction property and voltage-dependent state transitions from \cite{williams2013computational}. The irradiation intensity varied in the range of 0.165--0.175 mW/mm$^2$. 
 The forward Euler method was used to integrate the equations with a time step $\Delta t=0.02 $ ms  and a centered finite-differencing scheme to discretise the Laplacian with a space step of $\Delta x=0.0625  $ mm. The total computational domain size
was 256 $\times$ 256 grid points; while the centrally irradiated square with the optogenetic current consisted of 80 $\times$ 80 grid points. To create stationary initial conditions, the model was integrated for 2 minutes before a train of stimuli was applied from the upper border of the domain.

\begin{figure*}
\centering
\includegraphics[width=1.0\linewidth]{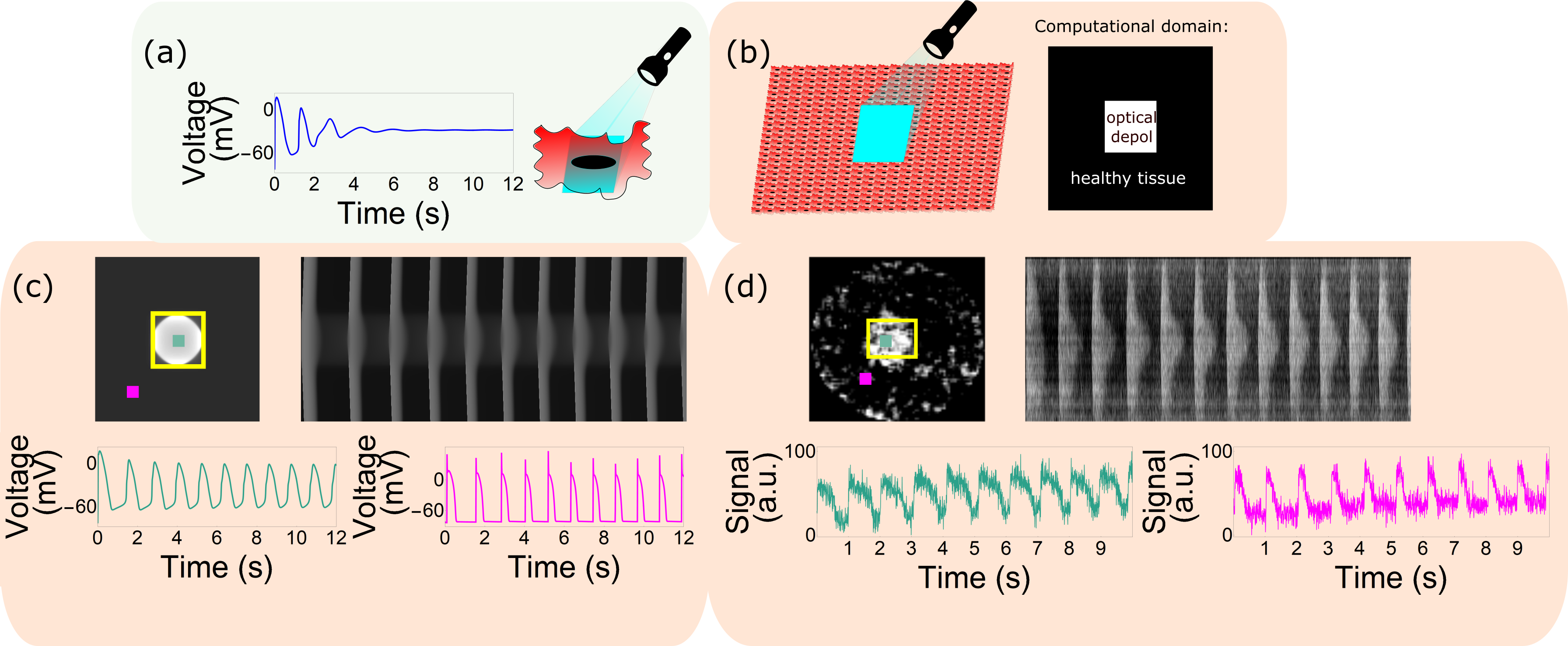}
\caption{ Sustained ectopic activity in experiment and computational model. (a) Single cell computational data for application of a optogenetic depolarizing tool, showing oscillations in transmembrane voltage until reaching a steady-state attractor. (b)  Scheme of experimental and numerical setup. (c) and (d) Sustained ectopy from the center of the illuminated area (yellow box) in computational model (c)  and  experiment (d).}
\label{fig:setup_model_exp}
\makeatletter
\let\save@currentlabel\@currentlabel
\edef\@currentlabel{\save@currentlabel(a)}\label{fig:setup_model_exp:a}
\edef\@currentlabel{\save@currentlabel(b)}\label{fig:setup_model_exp:b}
\edef\@currentlabel{\save@currentlabel(c)}\label{fig:setup_model_exp:c}
\edef\@currentlabel{\save@currentlabel(d)}\label{fig:setup_model_exp:d}
\makeatother
\end{figure*}

 Fig. \ref{fig:setup_model_exp}(a) shows the effect of prolonged optogenetic depolarization of the single cell, which is characterized by the transmembrane potential approaching a stable equilibrium upon light illumination with fixed intensity. 
\subsection{Optical mapping and patterned illumination of NRVM monolayers}
After 8-10 days of culturing, NRVM monolayers were optically mapped using the voltage-sensitive dye di-4-ANBDQBS (52.5 $\mu$M final concentration; ITK diagnostics, Uithoorn, the Netherlands) as reported previously \cite{feola2017localized}. The mapping setup was based on a 100 $\times$ 100 pixel CMOS Ultima-L camera  (Scimedia, Costa Mesa, CA). The field of view was 16 $\times$ 16 mm resulting in a spatial resolution of 160$\,\mathrm{\mu}$m/pixel. For targeted illumination of monolayers the setup was optically conjugated to a digitally controlled micro-mirror device (DMD), the Polygon 400 (Mightex Systems, Toronto, ON), with a high power blue  (470 nm) LED (BLS-LCS-0470-50-22-H, Mightex Systems). The scheme of the setup is shown in Fig. SI1C. Before starting the actual experiments, all monolayers were mapped during 1-Hz electrical point stimulation to check baseline conditions. Electrical stimulation was performed by applying 10-ms-long rectangular electrical pulses with an amplitude of 8 V to  a bipolar platinum electrode with a spacing of 1.5 mm between anode and cathode. Only cultures with an action potential duration (APD) at 80$\%$ repolarization (APD$_{80}$) below 350{ ms} and a CV above $18\,$cm/s were used for further experiments. The constant light intensity was varied in the range of  (0.03125-0.25\,mW/mm$^2$) to search the critical value for the bistable regime. For experiments in which the light intensity was varied the rectangular-shaped area of illumination was in the range of $4\times4$ to $6\times6$ mm. The highest achievable irradiation intensity (0.3125mW/mm$^2$) was used to perform size modulation experiments. The resulting electrical activity was recorded for 6-24 s at exposure times of 6 ms per frame.

\subsubsection{Postprocessing of optical mapping data}
Data processing was performed using specialized BV Ana software (Scimedia), ImageJ Ref.\cite{imagej} and custom-written scripts in Wolfram Mathematica (Wolfram Research, Hanborough, Oxfordshire, United Kingdom). APD and CV were calculated as described previously \cite{feola2017localized}. To prepare representative frames of wave propagation, optical mapping videos were filtered with a spatial averaging filter (3 $\times$ 3 stencil) and a derivative filter.
\subsubsection{Attactor reconstruction}
Multiple attractor reconstruction is based on Takens embedding theorem \cite{takens1981detecting} by plotting phase space trajectories in $(v(t),v(t+\tau))$ coordinates. The optimal time delay $\tau$ was determined by calculating the minimum of autocorrelation function \cite{clemson2014discerning}. Linear drift removal was applied before plotting phase space trajectories.

\section{Results}
\subsection{Optogenetically induced ectopic activity.}
\label{sec:general_setting}
We utilized 8 to 11-days-old monolayer of neonatal rat ventricular myocytes expressing the  depolarizing optogenetic tool  CheRiff as shown in Figs. SI1(a) and (b).
We found that  stationary ectopic activity can be  generated and maintained by  light-induced depolarization of a region within the monolayer [Fig. \ref{fig:setup_model_exp:b}]. 

 Fig. \ref{fig:setup_model_exp:d} shows sustained ectopic activity emerging from  the irradiated zone.    
 
Our experimental observations were corroborated with computational simulations using Majumder-Korhonen model of NRVM monolayers. As reported in the Methods section, our mathematical model also includes the description of light-gated ion channels with the rectification property of the channel conductance.
Fig. \ref{fig:setup_model_exp:c} shows periodic ectopic activity in the system after illumination of a small square at the center.
We also see that computational and experimental models produce similar excitation patterns.

Note, that in both examples  ectopy originates from the center of the area, which is being depolarized, and  not from the border of depolarized zone as in  \cite{teplenin2018paradoxical}. As we will see later, this is just a result of different degree of depolarization in the system induced by the optogenetic stimulus.  %

\begin{figure*}[]
\centering
\includegraphics[width=1.0\linewidth]{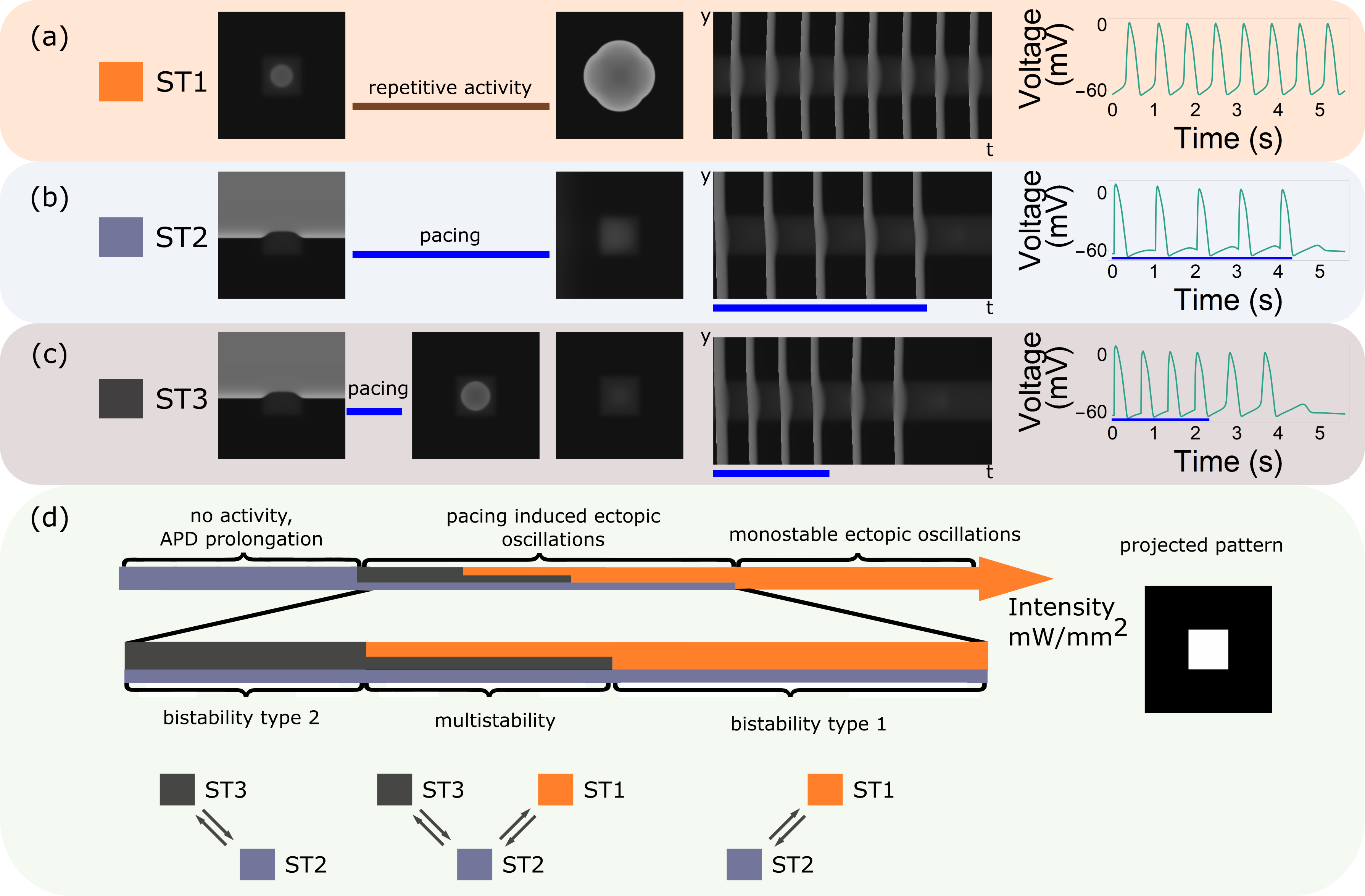}
\caption{ Bifurcation diagram.  (a)--(c) Observed dynamical states: (a)  sustained ectopy, state ST1; (b) quiescent  state before and after train of stimuli, state ST2; (c) transient ectopy after train of stimuli, state ST3. (d) Bifurcation diagram showing dependence of dynamical states on intensity of light (depolarizing current). }
\label{fig:bifurcation_diagram_2}
\makeatletter
\let\save@currentlabel\@currentlabel
\edef\@currentlabel{\save@currentlabel(a)}\label{fig:bifurcation_diagram_2:a}
\edef\@currentlabel{\save@currentlabel(b)}\label{fig:bifurcation_diagram_2:b}
\edef\@currentlabel{\save@currentlabel(c)}\label{fig:bifurcation_diagram_2:c}
\edef\@currentlabel{\save@currentlabel(d)}\label{fig:bifurcation_diagram_2:d}
\makeatother
\end{figure*}

\subsection{Bifurcation diagram}
\label{sec:bifurcation}
Next, we studied how the illumination intensity affects  ectopic activity. We first took a computational approach and then illustrated the main regimes in experimental studies.  Fig. \ref{fig:bifurcation_diagram_2} shows the observed changes in the excitation patterns represented as a qualitative  bifurcation diagram. We found that high-intensity irradiation produced sustained ectopic oscillations as shown in Fig. \ref{fig:bifurcation_diagram_2:a} and Figs. \ref{fig:setup_model_exp:c}, \ref{fig:setup_model_exp:d}.  Note that if irradiation is very intense the activity comes from the corners of the irradiated domain (as in \cite{teplenin2018paradoxical} paper, not shown). For lower irradiation intensities, it shifts and starts form the geometrical center of the irradiated zone as in Fig. \ref{fig:bifurcation_diagram_2:a}.  We  denote the stable sustained  oscillatory regime as  state 1 (ST1). After reduction of the irradiation intensity, the ectopic waves were no longer induced after the start of irradiation. The media therefore remained in a state showing stationary spatial distribution of the voltage. This quiescent state is denoted by state 2 (ST2). However, we found that under some  illumination conditions the system  can be perturbed by travelling excitation waves as shown in  Fig. \ref{fig:bifurcation_diagram_2:b}. In particular, we observed that if the wave train propagated through the whole media and, consequently, through the illuminated region, sustained ectopic oscillations were subsequently evoked from this region. 

Thus, such invading waves induced transition from state ST2 to ST1, thereby unmasking the bistable nature of the system in this parametric range. We have designate this  bistability between ST1 and ST2 as bistability of type 1 [Fig. \ref{fig:bifurcation_diagram_2:d}]. 

If the irradiation intensity was decreased further, the sustained ectopic waves were no longer induced, and after the start of irradiation  the system always resided in ST2 for any initial conditions and perturbations. We also found that for some range of intensities the system could produce  transient ectopic oscillations. Those spatio-temporal oscillations were similar to oscillations of ST1 in location and shape, but had a transient nature.  It was observed [Fig. \ref{fig:bifurcation_diagram_2:c}] that for this parametric range the system could be perturbed by a train of external waves, and after such stimulation (end of the blue line below the right sub-figure) there were two additional waves emerging from the heterogeneity. We denoted such transient activity as state 3 (ST3). Strictly speaking, ST3 could not be called a state in a strict mathematical sense, since the system finally goes to the resting state ST2. However,  in our view, we can still call  ST3 a state or pseudostate, because transient ectopic activity or extrasystoles have  clinical significance as they are the main   trigger of cardiac  arrhythmias. Therefore, considering ST3 as a separate pseudostate has also  practical importance. With state ST3 we can introduce additional bifurcation regimes.  We also found that in some parametric range external waves, depending on their properties, cause either transient ectopic activity (ST3), or sustained ectopic activity (ST1) or no ectopic activity at all (ST2). Thus we have a region of multistability [Fig. \ref{fig:bifurcation_diagram_2:d}]. Also below the values of illumination for establishing ST1, we could find coexistence of ST2 and ST3, which was denoted as bistability of type 2 [Fig. \ref{fig:bifurcation_diagram_2:d}]. Obviously, a bistability or multistability involving ST3 is not correct in a strict mathematical sense, as ST3 eventually evolves to ST2. However, we will keep our terminology through the paper, due to practical importance of regime ST3.

Interestingly, transitions $ST2\xrightarrow{}ST1$  and $ST2\xrightarrow{}ST3$  showed particular resonance properties, which are described in the next section.

\begin{figure*}[]
\centering
\includegraphics[width=1.0\linewidth]{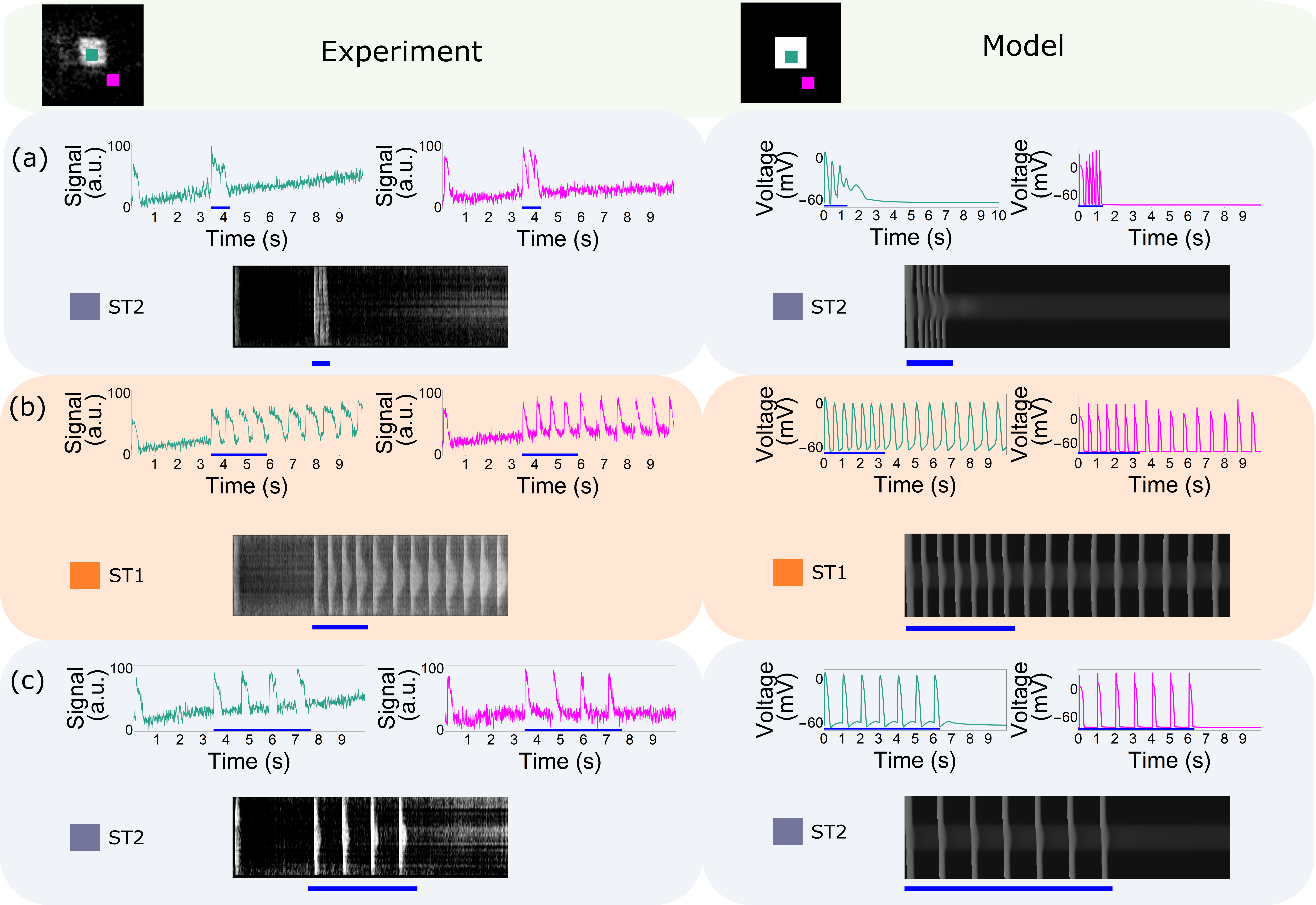}
\caption{ Resonant induction of ectopic activity \textit{in-vitro} (left panels) and \textit{in-silico} (right panels). The time window for stimulation is indicated by blue line. (a) High frequency stimulation experiment does not induce ectopic activity. (b) Medium frequency stimulation does induce ectopic activity, while (c) low frequency stimulation fails to induce ectopic activity.  }
\label{fig:resonance_switch_on}
\makeatletter
\let\save@currentlabel\@currentlabel
\edef\@currentlabel{\save@currentlabel(a)}\label{fig:resonance_switch_on:a}
\edef\@currentlabel{\save@currentlabel(b)}\label{fig:resonance_switch_on:b}
\edef\@currentlabel{\save@currentlabel(c)}\label{fig:resonance_switch_on:c}

\makeatother
\end{figure*}

\begin{figure*}[]
\centering
\includegraphics[width=1.0\linewidth]{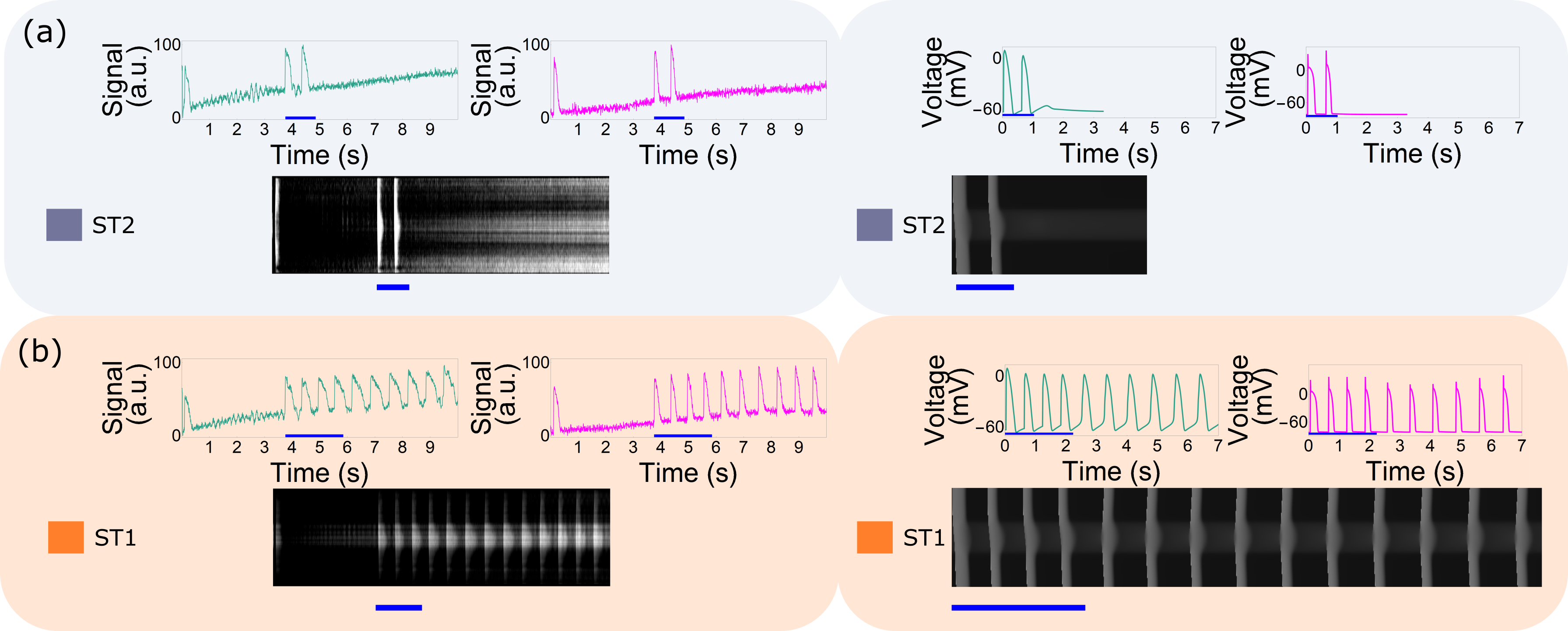}
\caption{ Minimal number of travelling waves to induce ectopic activity \textit{in-vitro} (left panels) and \textit{in-silico} (right panels). (a) Two pulses of 600ms are not enough to initiate ectopy. (b) Four pulses of 600ms are sufficient to initiate ectopic activity.}
\label{fig:2_vs_4}
\makeatletter
\let\save@currentlabel\@currentlabel
\edef\@currentlabel{\save@currentlabel(a)}\label{fig:2_vs_4:a}
\edef\@currentlabel{\save@currentlabel(b)}\label{fig:2_vs_4:b}
\makeatother
\end{figure*}

\begin{figure}[]
\centering
\includegraphics[width=1.0\linewidth]{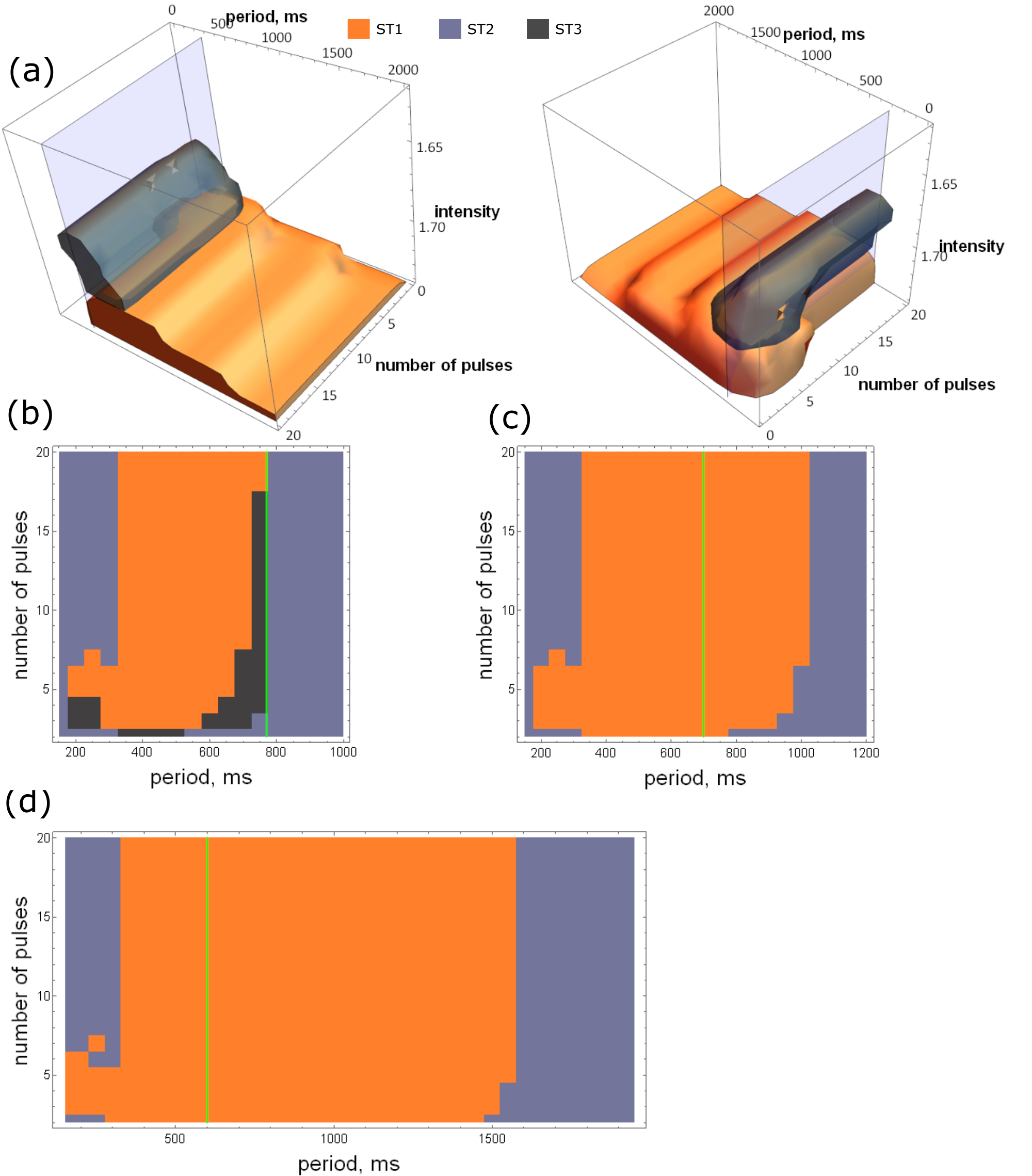}
\caption{ 3-parametric state diagram.  (a) Two different angle views of 3D diagram of 3 parameters: pulse number, pulse period andlight  intensity. (b)--(d) Sections of 3D diagram for intensity 1.7124, 1.72 and 1.73 respectively, the vertical green lines indicate the period of sustained ectopy. (b) Multistability between ST1, ST2 and ST3. (c) and (d) Bistability between ST1 and ST2. }
\label{fig:3D}
\makeatletter
\let\save@currentlabel\@currentlabel
\edef\@currentlabel{\save@currentlabel(a)}\label{fig:3D:a}
\edef\@currentlabel{\save@currentlabel(b)}\label{fig:3D:b}
\edef\@currentlabel{\save@currentlabel(c)}\label{fig:3D:c}
\edef\@currentlabel{\save@currentlabel(d)}\label{fig:3D:d}
\makeatother
\end{figure}

\subsection{Resonant transitions  from quiescence to state ST2}
\label{sec:decription_resonance}

Resonance properties of the transition $ST2\xrightarrow{}ST1$ are illustrated in  Fig. \ref{fig:resonance_switch_on}, which shows  side-by-side experimental (left column) and modelling (right column) results. In all cases, we initiated 4 travelling waves from the border of the domain with various periods. The stimulation time is marked by the blue stripe.
We see that for low stimulation period, which is close to the refractory period of the tissue (typically 180--350 ms in the \textit{in-vitro} experiment and 150--250 ms in the computational modelling)  no sustained ectopic activity occurs [Fig. \ref{fig:resonance_switch_on:a}].  Instead, we observed tissue going to the steady state ST2 directly  after the end of the stimulation.  Note, that in the experiments, there are a few seconds of quiescence before the stimulation (the blue line) demonstrating the absence of ectopic activity before the start of the  stimulation. Next,  we observed that intermediate stimulation periods (400--800 ms in the \textit{in-vitro} experiment and 300--1000 ms in the computational modelling) induced sustained ectopic activity [Fig. \ref{fig:resonance_switch_on:b}]. Finally, using longer stimulation period ($>800$ ms in the \textit{in-vitro} experiment and $>1000$ ms in the computational modelling)  we again observed no initiation of sustained activity [Fig. \ref{fig:resonance_switch_on:c}].  Thus we can conclude that initiation of the ectopic activity has resonance properties. 
We found that induction of the ectopic activity also depends on the number of external waves. Fig. \ref{fig:2_vs_4}
 shows experimental and computational results for initiation of ectopic activity by 2 or 4 waves with the same period. We see that ectopic activity is induced after 4 waves [Fig. \ref{fig:2_vs_4:b}] but not after 2 waves [Fig. \ref{fig:2_vs_4:a}]. This means we also need a sufficient number of waves to initiate the ectopic activity.    
 
\begin{figure*}[]
\centering
\includegraphics[width=1.0\linewidth]{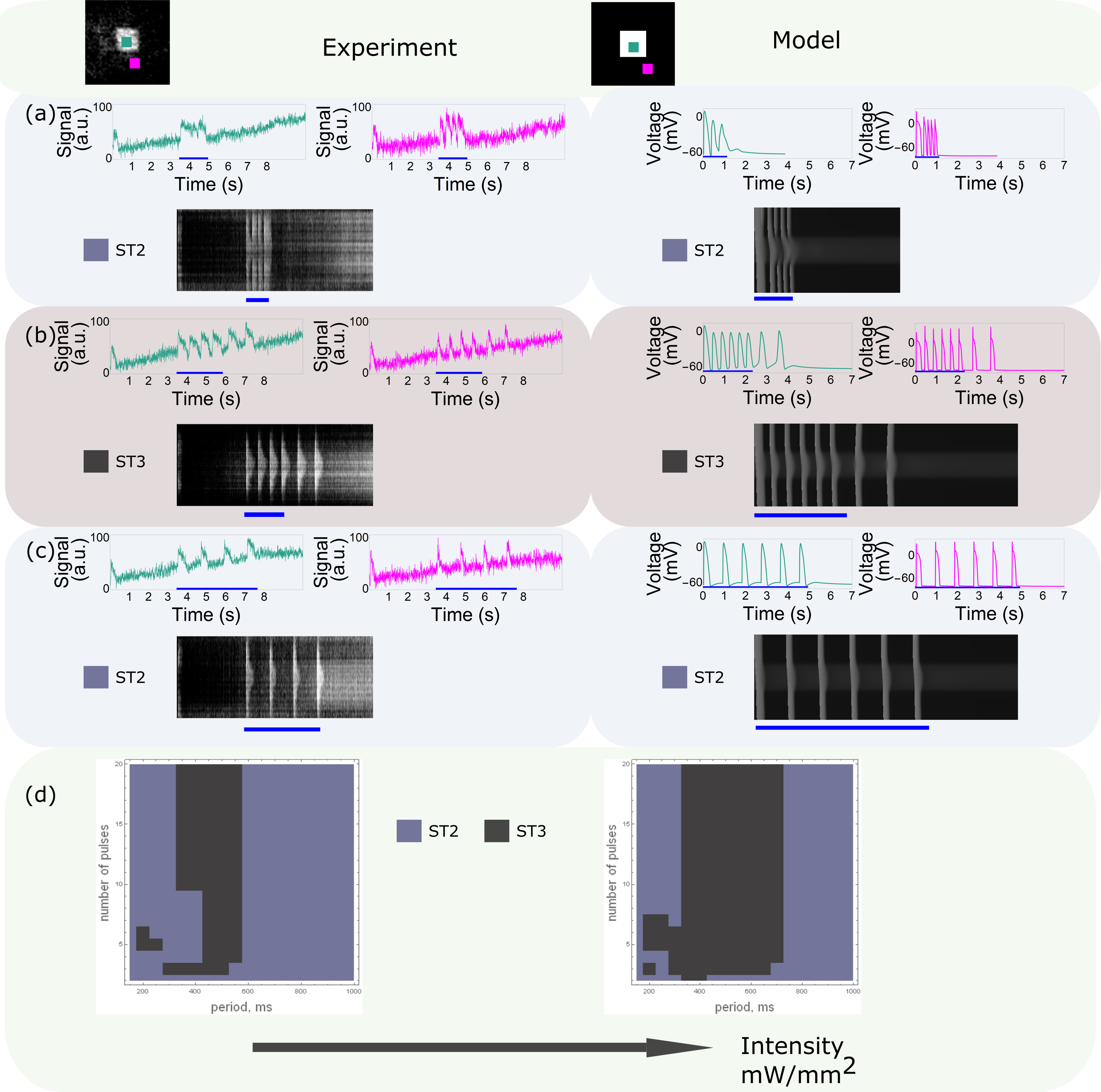}
\caption{  Resonance for initiation of transient activity. Periodic forcing with low (a), intermediate (b) and high (c) periods \textit{in-vitro} (left panels) and \textit{in-silico} (right panels). (d) Section of 3D diagram  from 3D graph in Fig. \ref{fig:3D} showing bistability between ST1 and ST2 for intensities 1.7 (left panel) and 1.71 (right panel) correspondingly.}
\label{fig:transient}
\makeatletter
\let\save@currentlabel\@currentlabel
\edef\@currentlabel{\save@currentlabel(a)}\label{fig:transient:a}
\edef\@currentlabel{\save@currentlabel(b)}\label{fig:transient:b}
\edef\@currentlabel{\save@currentlabel(c)}\label{fig:transient:c}
\edef\@currentlabel{\save@currentlabel(d)}\label{fig:transient:d}

\makeatother
\end{figure*}

  All these phenomena, including bistability itself, also  depend on the intensity of the illumination [Fig. \ref{fig:bifurcation_diagram_2:d}]. Thus overall, initiation of ectopy in the bistable regime should depend on three parameters: period and number of waves and light intensity.  Since it is not feasible to perform a full three-parametric-study \textit{in-vitro} these studies were performed \textit{in-silico}. The results of these studies are shown in Fig. \ref{fig:3D}. A full three-parametric diagram is shown in Fig. \ref{fig:3D:a} from two different angle views. The light intensity is represented by the $z$-axis and increases downwards.  We noticed that the range of periods initiating  sustained ectopic activity (depicted by the orange colour) increases with increasing light intensity. Figs. \ref{fig:3D:b}, \ref{fig:3D:c} and \ref{fig:3D:d} illustrate this in 2D sections of the graph for three different values of the intensity. We also observed that the number of pulses necessary to induce ectopic activity decreases with the increasing light intensity. For example, for a simulation period of 1000 ms and the intensity  shown in Fig. \ref{fig:3D:b}, no ectopic activity could be induced by any number of pulses, whereas for the intensities in Figs. \ref{fig:3D:c} and \ref{fig:3D:d} ectopic activity could be induced by 7 and 2 pulses, respectively. The transient oscillations regime (ST3) occurred only at the low intensity, which means that here we indeed have multistability between states ST1, ST2 and ST3. The frequency of sustained ectopic activity is represented by the green line in Figs. \ref{fig:3D:b}, \ref{fig:3D:c}, \ref{fig:3D:d}, showing a decrease with increasing light intensity.  Also note that the green line is at the right boundary of the resonant period range for the low light intensity, while it is at the middle for the intermediate and at the left for the high light intensity. Thus resonant range is not directly defined by the period of self oscillations.
 \newline Let us also consider in more details the phenomena observed for the lower light intensity. If the light intensity is sufficiently low, then we do not have sustained oscillations (ST2) and observe only transient  ectopic activity (ST3).  This situation is depicted in Fig. \ref{fig:transient} where we  present  side-by-side both experimental and modelling results. As in Fig. \ref{fig:resonance_switch_on}, we found initiation of ectopy by external waves with different periods. We observed that only waves with the intermediate period induced transient oscillations [Fig. \ref{fig:transient:b}], i.e. waves with lower and higher periods did not induce transient oscillations [Fig. \ref{fig:transient:a}, \ref{fig:transient:c}]. Thus, here we have a resonance phenomenon for bistability of type 2 displaying similar resonance properties as bistability of type 1 (see above). For example, the range of resonant periods increased with increasing intensity [Fig. \ref{fig:transient:d}]. We also noticed the experimental results  closely followed the theoretical findings. 

\subsection{Termination of ectopic activity}
\label{sec:termination}
\begin{figure*}
\centering
\includegraphics[width=1.0\linewidth]{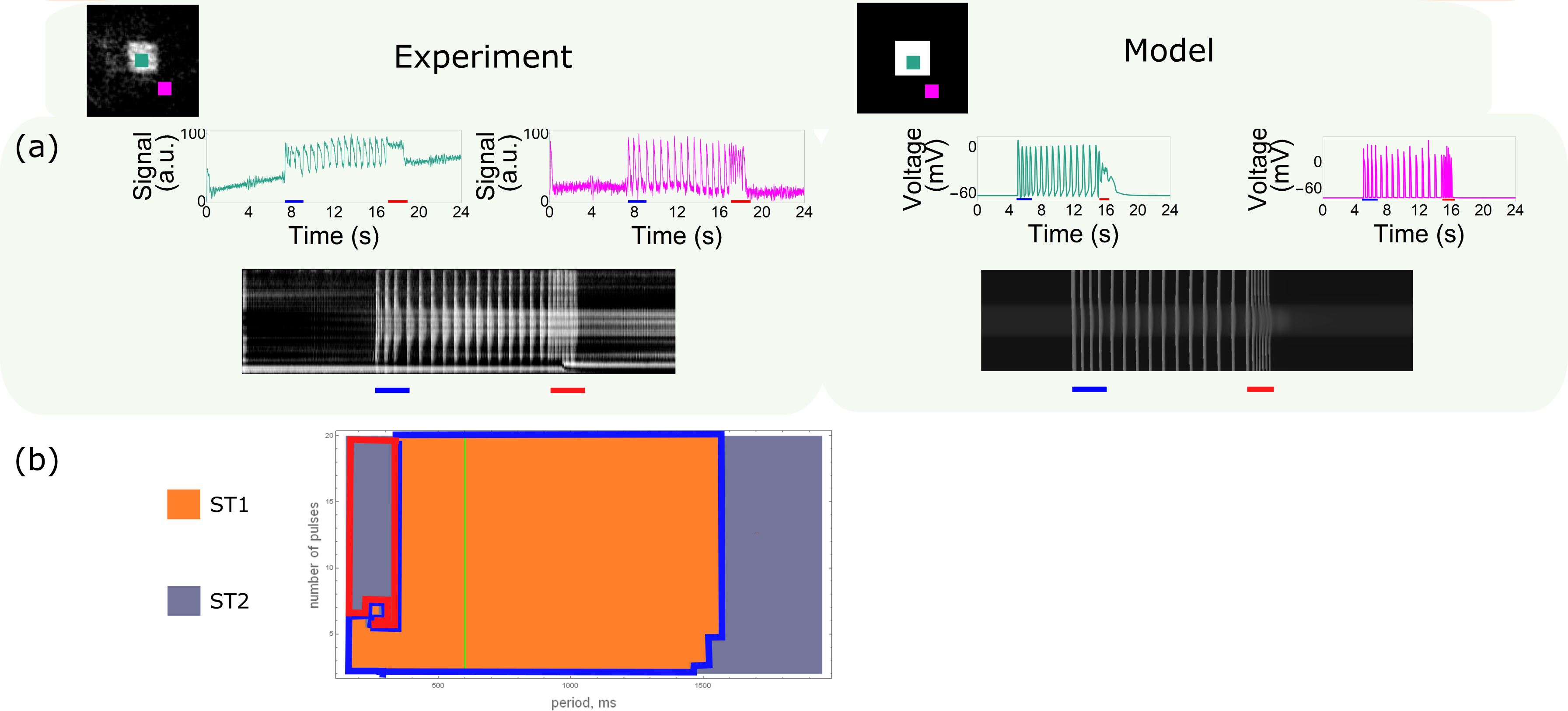}
\caption{ (a) Initiation and termination of ectopic activity \textit{in-vitro} (left panel) and \textit{in-silico} (right panel). (b) Section of  diagram  from 3D graph in Fig. \ref{fig:3D} for intensity 1.73 with indicated termination (red) and initiation (blue) period ranges. }
\label{fig:termination}
\makeatletter
\let\save@currentlabel\@currentlabel
\edef\@currentlabel{\save@currentlabel(a)}\label{fig:termination:a}
\edef\@currentlabel{\save@currentlabel(b)}\label{fig:termination:b}
\makeatother
\end{figure*}

 Another interesting observation  from Fig. \ref{fig:3D} is that sustained ectopic activity can potentially be stopped by external wave stimulation. In Fig. \ref{fig:3D} we found the conditions for initiation of ectopic activity from quiescent state (ST2).  Now let us consider a  system with sustained ectopic activity (ST1) and let us perturb it by external waves with a given period.  
 In many cases external stimulation  will have no effect.  Indeed, if the period of the external waves is longer than the period of self-oscillations [the green line in Fig. \ref{fig:3D:d}], then the external waves will not reach the ectopic region and thus will not influence it. If, however, the ectopic region is stimulated with a shorter period than the self-oscillation period the external source will overdrive the ectopic source and external waves will be able to propagate through the region, effectively resulting in a situation similar to that studied in Fig. \ref{fig:3D} for high frequency of stimulation. If the conclusions of Fig. \ref{fig:3D} hold for this case,  external waves with the period  in the violet parametric region of Fig. \ref{fig:3D}  should stop the sustained ectopic activity. 
 We performed studies of such situation \textit{in-vitro} and \textit{in-silico} [Fig. \ref{fig:termination:a}]. We first initiated sustained ectopic source by external stimulation (the blue line) and after some while applied an additional train of traveling waves with the period of the left violet region in Fig. \ref{fig:termination:b} (encircled by the red line). This resulted in termination of the ectopic activity after the ending of the high frequency waves. Thus we indeed can not only induce, but we can also terminate the sustained ectopic activity by external high frequency waves. Once again the experimental results closely followed theoretical findings.

\subsection{Collective nature of oscillations and resonance}
\label{sec:collective_nature}

\begin{figure*}[]
\centering
\includegraphics[width=1.0\linewidth]{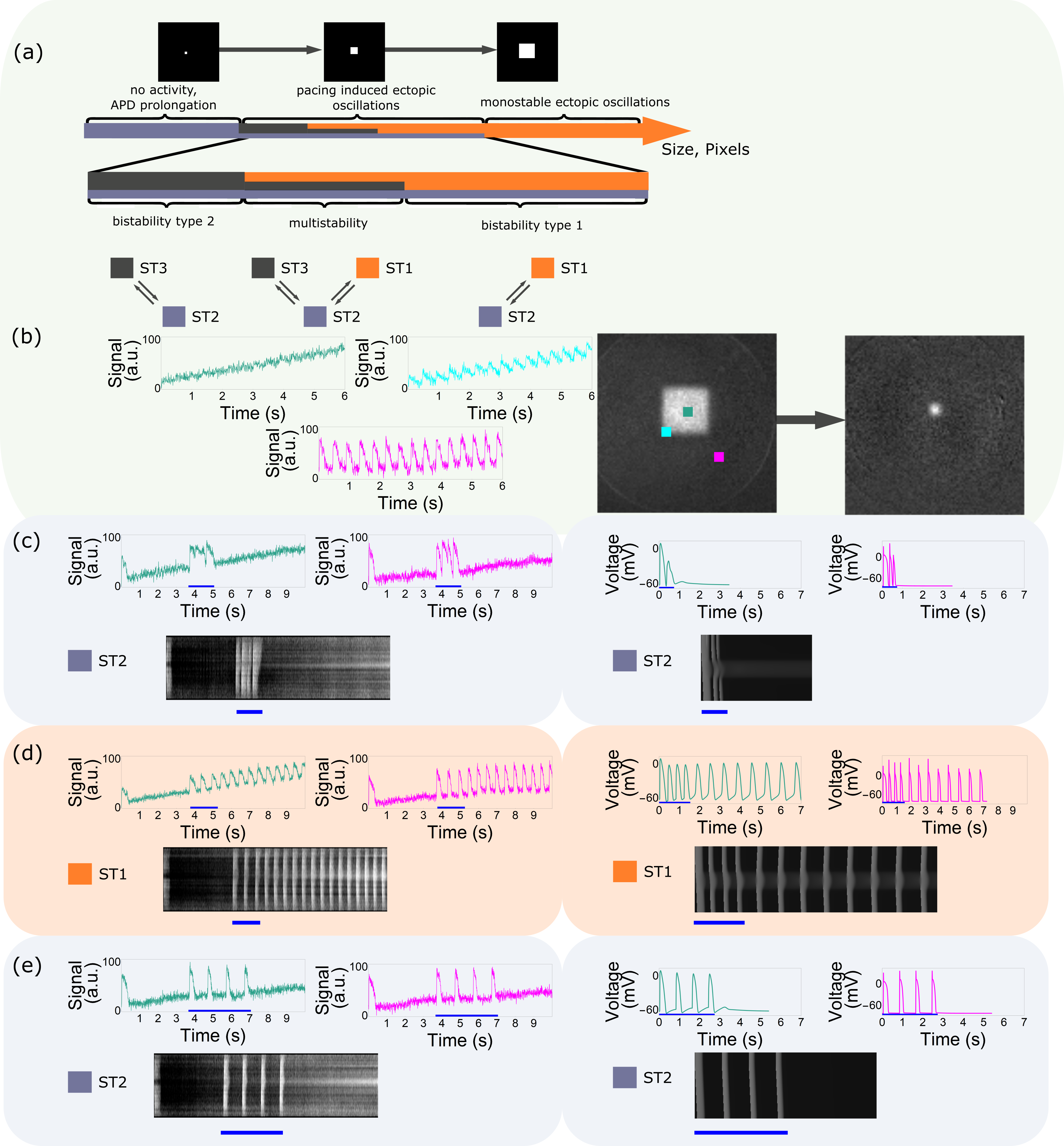}
\caption{ Effect of size of the depolarized region. (a) Bifurcation diagram with the size of the region as parameter. (b) Left part: example of ectopic activity from a large illuminated area with no oscillations at the center of the region. Right part: example of small region, which allows existence of bistability regime, illuminated with the same intensity. (c)-(e) Resonant induction of ectopic activity for the small illuminated area shown in (b). (c) High frequency stimulation experiment does not induce ectopic activity. (d) Medium frequency stimulation does induce ectopic activity, while (e) low frequency stimulation fails to induce ectopic activity. }
\label{fig:size_resonance}
\makeatletter
\let\save@currentlabel\@currentlabel
\edef\@currentlabel{\save@currentlabel(a)}\label{fig:size_resonance:a}
\edef\@currentlabel{\save@currentlabel(b)}\label{fig:size_resonance:b}
\edef\@currentlabel{\save@currentlabel(c)}\label{fig:size_resonance:c}
\edef\@currentlabel{\save@currentlabel(d)}\label{fig:size_resonance:d}
\edef\@currentlabel{\save@currentlabel(d)}\label{fig:size_resonance:e}
\makeatother
\end{figure*}

Both normal and ectopic electrical activity in the heart are the result of  large amplitude oscillatory dynamics occurring at a single cell level. It is also known that one can   observe  bistability at the single cell level \cite{guevara2003bifurcations}.
However,  in our case the situation is different as we have a purely collective effect: large amplitude oscillatory dynamics and bistability in our system are not present at the single cell level as shown \textit{in-silico} in Fig. \ref{fig:setup_model_exp:a} for  the same light intensity as the one causing sustained ectopic activity in  Fig. \ref{fig:setup_model_exp:c}. Unfortunately, it is not feasible to perform a similar experiment \textit{in-vitro} would require measuring the electrical activity of a single cell within the illuminated region under unrealistic  conditions, meaning that electrotonic effects of  non-illuminated areas would be absent. One way to diminish electrotonic effects of non-illuminated  areas is to increase the size of the illuminated region.  
Fig. \ref{fig:size_resonance:b} shows oscillatory ectopic activity (ST1) for this case. We clearly see  ectopic waves originating from the border of the illuminated region (magenta trace) but do not observe oscillations at the center (green trace) . This indeed indicates that single cell dynamics in our system upon illumination is not oscillatory, but that such oscillation is a collective phenomenon. Note that the monotonic increase of the green and magenta signals is the baseline drift due to renormalization of the optical signal and thus does not represent any gradual change of voltage.

\begin{figure}[]
\centering
\includegraphics[width=1.0\linewidth]{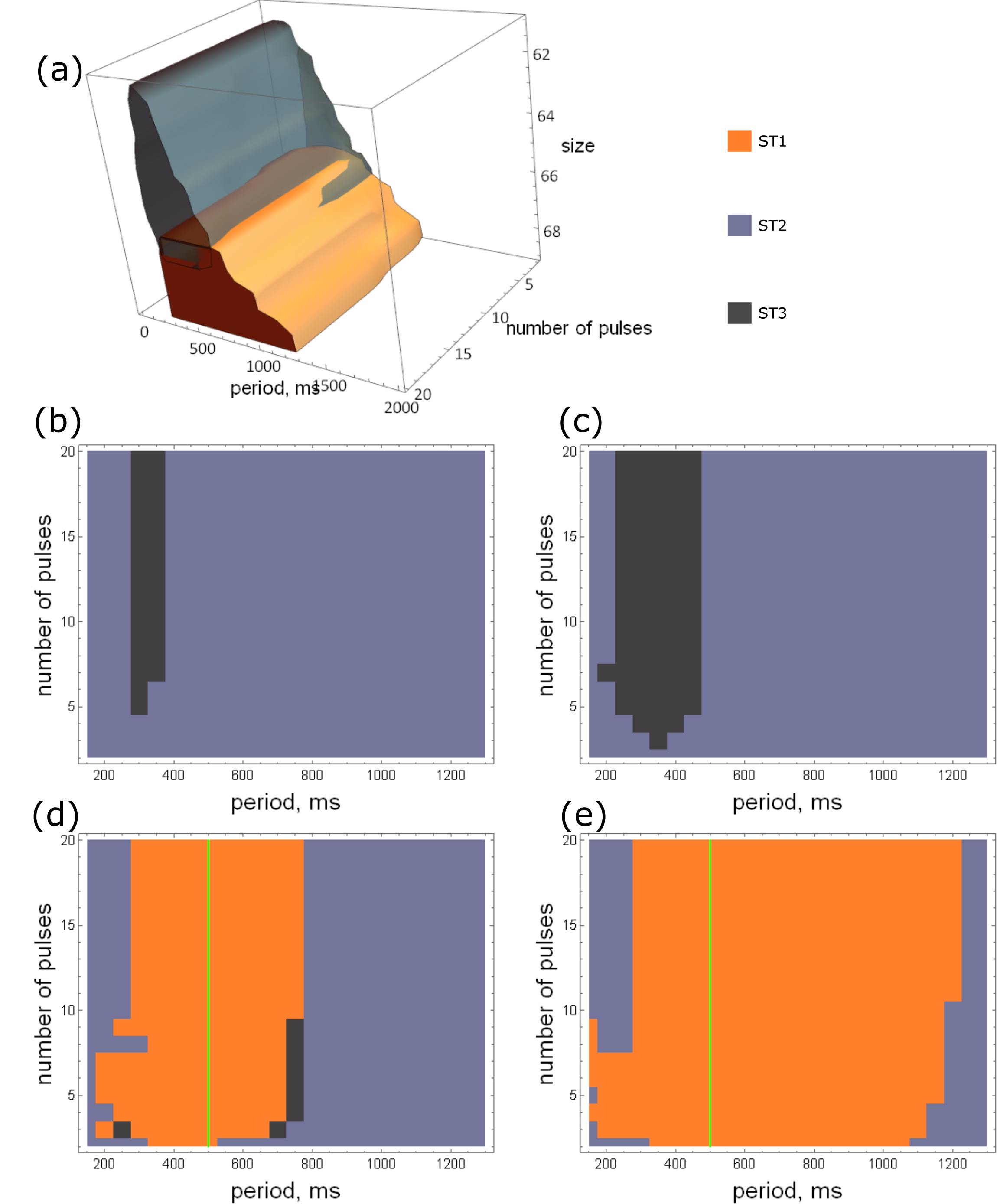}
\caption{ Diagrams for different sizes of areas of illumination at a constant intensity . (a) 3D diagram of 3 parameters:  number of pulses, pulse period and size of the illuminated region. (b)--(e) Sections of 3D diagram for pattern of 3.875$\times3.875$, 4$\times4$, $4.1875\times4.1875$ and 4.3125$\times$4.3125 mm$^2$ respectively. The vertical green lines indicate the period of sustained ectopy. (b) and (c) show bistability between ST2 and ST3. (d) Multistability between ST1, ST2 and ST3. (e) Bistability between ST1 and ST2.}
\label{fig:small_3D}
\makeatletter
\let\save@currentlabel\@currentlabel
\edef\@currentlabel{\save@currentlabel(a)}\label{fig:small_3D:a}
\edef\@currentlabel{\save@currentlabel(b)}\label{fig:small_3D:b}
\edef\@currentlabel{\save@currentlabel(c)}\label{fig:small_3D:c}
\edef\@currentlabel{\save@currentlabel(d)}\label{fig:small_3D:d}
\edef\@currentlabel{\save@currentlabel(e)}\label{fig:small_3D:e}
\makeatother
\end{figure}

Next, while maintaining a constant light intensity, we decreased the size of region and found the bifurcation diagram depicted in Fig.  \ref{fig:size_resonance:a}. This bifurcation diagram has a similar structure as the diagram in Fig. \ref{fig:resonance_switch_on} for varying light intensity and also shows bistability and resonance properties. Bistability can be clearly seen in Fig.   \ref{fig:size_resonance:a}. To achieve bistability we decreased the size of the illuminated area from 7$\times$7 mm$^2$ to 1$\times$1 mm$^2$ as illustrated in the right half of Fig. \ref{fig:size_resonance:b}. This bistability also had resonant properties as shown in Figs. \ref{fig:size_resonance:c}, \ref{fig:size_resonance:d}, \ref{fig:size_resonance:e}. Similar to findings in Fig. \ref{fig:resonance_switch_on}, waves of short and long period did not induce ectopic activity \textit{in-vitro} [left parts of Fig. \ref{fig:size_resonance:c} and Fig. \ref{fig:size_resonance:e}, respectively], while waves of an intermediate period triggered stable oscillations [left part of Fig. \ref{fig:size_resonance:d}]. Thus the resonance observed with a decrease of the size of the illuminated region is of the same kind as with a decrease of light intensity. 
Note that for the illuminated area of 1 mm,  oscillations penetrate to the center due to electrotonic effects. However,  we know that here the single cell dynamics is non-oscillatory. 
The \textit{in-vitro} results could be also reproduced \text{in-silico} [right parts of Figs. \ref{fig:size_resonance:c}, \ref{fig:size_resonance:d}, \ref{fig:size_resonance:e}], which allowed us to generate the 3D parametric bifurcation diagram shown in Fig. \ref{fig:small_3D}. This bifurcation diagram displays similar properties as the bifurcation diagram in Fig. \ref{fig:3D}, such as a widening of the range of stimulation periods capable of initiating  ectopic activity. Such widening occurs with an increase in the size of the illuminated area [see Figs. \ref{fig:small_3D:b}, \ref{fig:small_3D:c}, \ref{fig:small_3D:d}, \ref{fig:small_3D:e}] analogous to the effect of an increase in light intensity as shown in Figs. \ref{fig:3D:b}, \ref{fig:3D:c}, \ref{fig:3D:d}.
\section{Mechanism of the phenomenon}
\label{sec:mechanism}
To elucidate the mechanism of resonance as observed in our study it is helpful to compare our system to other nonlinear biological systems with frequency selection properties and apply the mathematical frameworks used to study these properties. In biological systems such as neurons,  frequency selection is usually associated with type II excitability governed by so-called membrane resonance at a single cell level, or mathematically by damped or non-damped subthreshold oscillations initiated due to proximity of its resting state to a Hopf bifurcation \cite{hutcheon2000resonance,IZHIKEVICH2000}.   However, this mechanism does not apply to  our phenomenon. Firstly, we do not observe the bands of resonant periods in our bifurcation diagram. On the contrary, in classical resonant mechanisms the system should respond to all periods, which are integer ratios of the minimum resonant period. Secondly, we do not observe any subthreshold oscillations around the resting state. Below we provide additional arguments to exclude  Hopf bifurcation as the mechanism of the dynamics observed in our paper and outline an explanation for the  frequency selection phenomena. 
\begin{figure}
\centering
\includegraphics[width=1.0\linewidth]{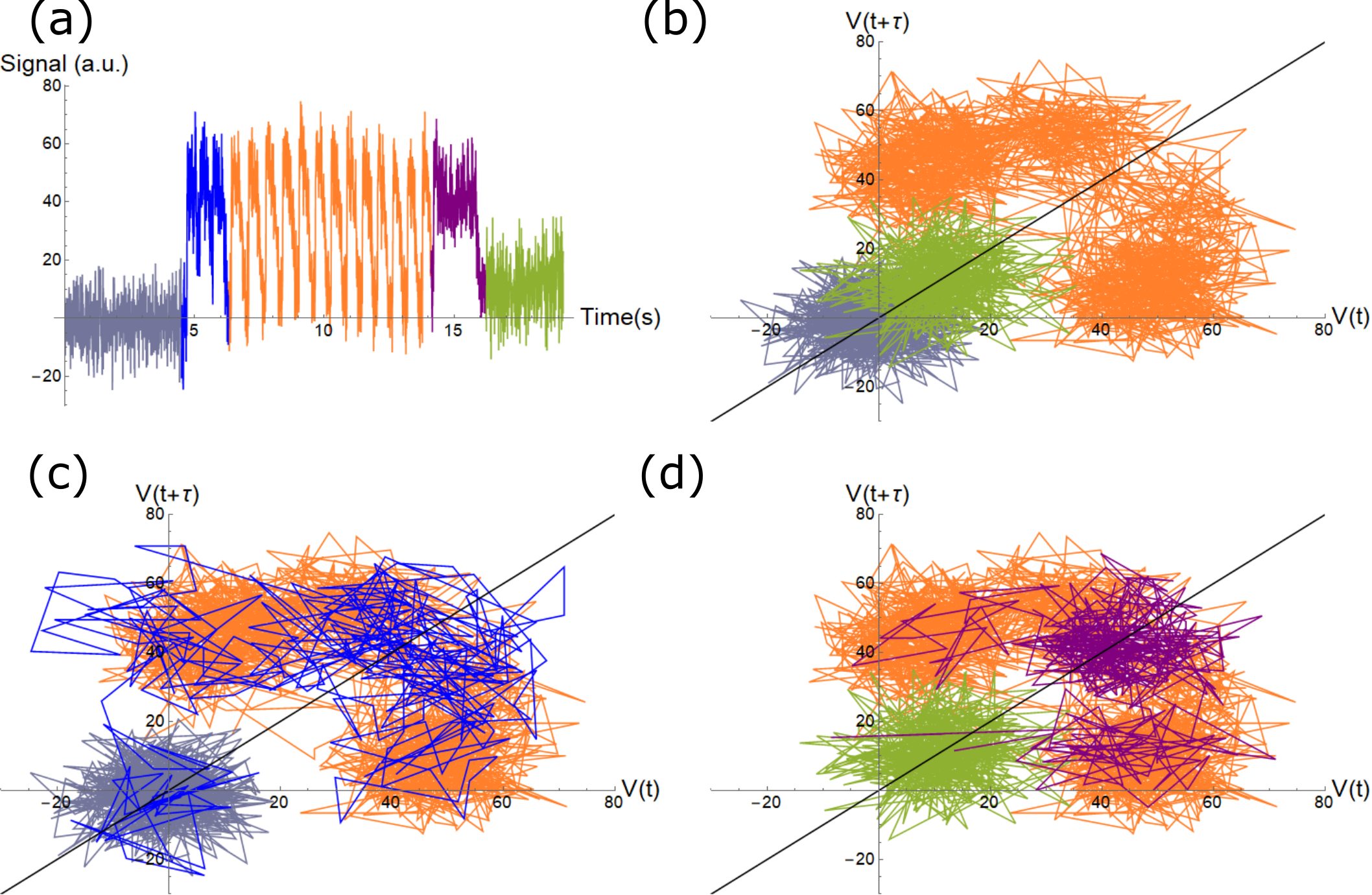}
\caption{Reconstruction of trajectories corresponding to observed dynamics in phase space using Takens time delay embedding. (a) The optical voltage signal showing induced initiation and termination of the ectopic activity. Optical trace was used for reconstruction of the signal. The initial non-depolarized state is shown in grey, the initiating stimulating pulses in blue, periodic oscillatory attractor in orange, the terminating high frequency stimulation in purple, non-depolarized state after termination of oscillatory activity in light green. The green and grey trace should coincide in intensity level but differ due to drift of the signal. (b), (c), (d) Trajectories in phase space in $(v(t),v(t+\tau))$ coordinates. Three separate plots are shown to avoid overlapping of the trajectories. b) Multiple attractors. Orange area --- oscillatory (i.e. ST1) attractor. Grey and green --- ST2 attractor. (c) Approaching   ST1 attractor by  medium frequency stimulation (blue trajectory). (d) Departing from ST1 attractor by high frequency stimulation (purple trajectory). }
\label{fig:explanation_threshold_exp}
\makeatletter
\let\save@currentlabel\@currentlabel
\edef\@currentlabel{\save@currentlabel(a)}\label{fig:explanation_threshold_exp:a}
\edef\@currentlabel{\save@currentlabel(b)}\label{fig:explanation_threshold_exp:b}
\edef\@currentlabel{\save@currentlabel(c)}\label{fig:explanation_threshold_exp:c}
\edef\@currentlabel{\save@currentlabel(d)}\label{fig:explanation_threshold_exp:d}

\makeatother
\end{figure}
\subsection{Two dynamical attractors}
\label{sec:two_attr}

 An important property of bistability involving Hopf bifurcation is that the resting state is located inside the limit cycle.  However, in our case the resting state is stable and does not lie inside the limit cycle. In Fig. SI2 we show 2D projections of phase space trajectories corresponding to sustained ectopic activity (state ST1) and to a steady  state ST2 in computer simulations.  We see that in all 6 presented projections the steady state is outside the limit cycle. We did similar studies for the experimental data [Fig. \ref{fig:explanation_threshold_exp:a}]. Here we use time-delayed embedding \cite{takens1981detecting} to show all important types of dynamics. It starts from steady state ST1 (the grey color), then we apply external waves (blue color), followed by sustained activity (orange color), and overdrive pacing (violet  color) to finally reach steady state (green color). We also see that steady state ST2 is clearly located outside the limit cycle. This relative position indicates the existence of threshold manifold (separatrix) between the resting state and limit cycle. Importantly, the threshold manifold is not located in the area inside the limit cycle. That explains another property of our system, namely that it requires several external waves to initiate sustained ectopic activity. This is because several pulses accumulate to reach certain values of voltage and state variable, thereby allowing to cross the threshold manifold and land in the basin of attraction of the limit cycle. This is similar to the behaviour of an integrate-and-fire neuron \cite{IZHIKEVICH2000}, where multiple pulses are required to accumulate in order to reach the voltage threshold. However, our case is more complex, since the threshold manifold should be determined not only by a single level of voltage but by other state variables as well.

\subsection{"Hidden" bistability and perturbations of different magnitude}
\label{sec:hidden_bist}
\begin{figure}
\centering
\includegraphics[width=1.0\linewidth]{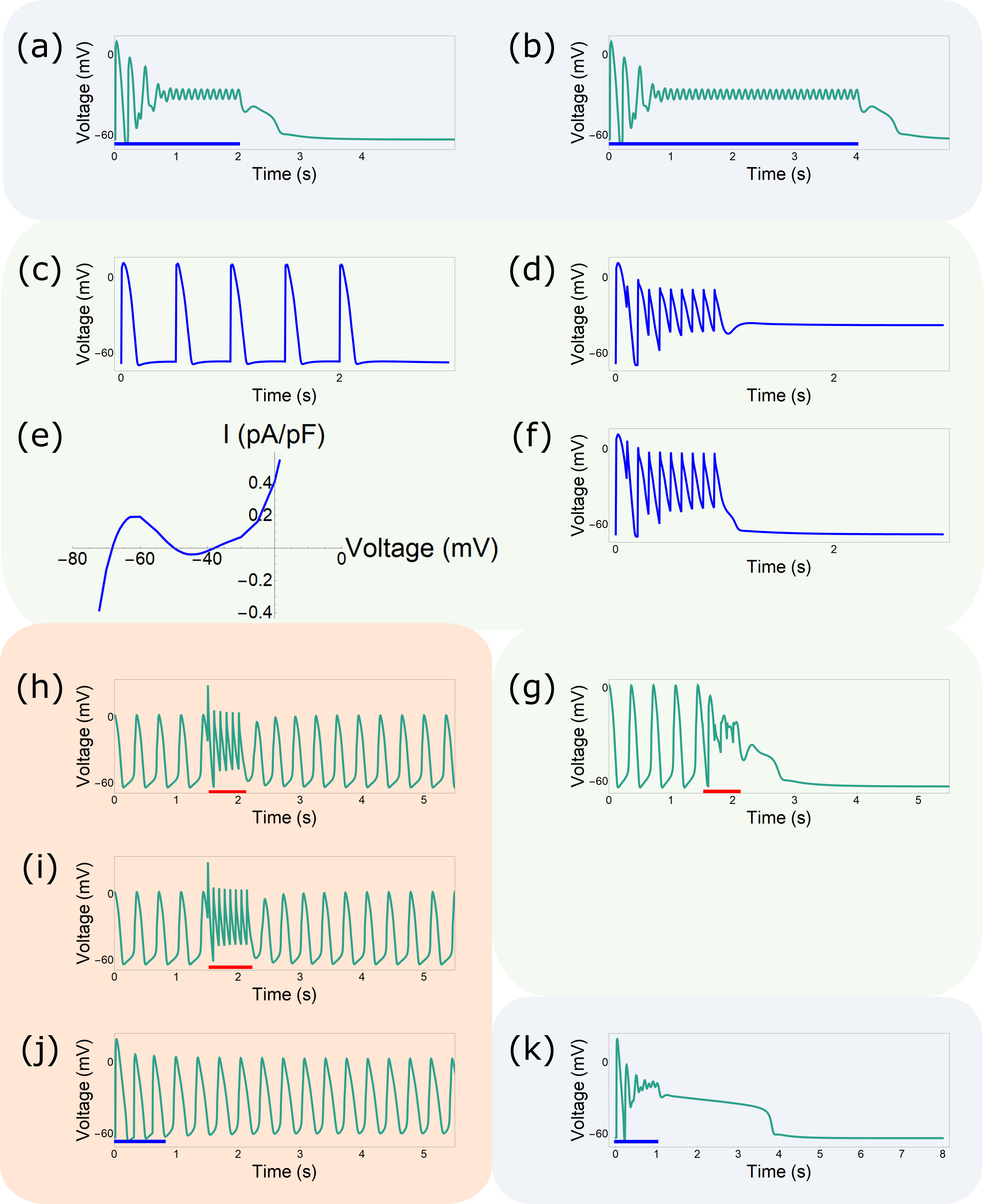}
\caption{ Formation of the quasi steady-state in the center of irradiated region during a train of 20 (a) and 40 (b) high frequency (180 ms) pulses. (c)  Excitable dynamics for low frequency of stimulation with a magnitude of $20$ mV of voltage perturbations. (d) Transition to the depolarized state after a train of high frequency stimuli with a magnitude of $20$ mV. (e) N-shaped  IV curve of whole-cell steady-state current. (f)  Excitable dynamics for a high frequency of stimulation and high magnitude of $50$ mV  in contrast to situation in (d). (h) Failure to terminate ectopic activity by spatially uniform voltage perturbations of high frequency and high magnitude ($50$ mV). (g) Termination of ectopic activity by spatially uniform voltage perturbations of high frequency and intermediate magnitude ($0$ mV). (i) Failure to terminate ectopic activity by spatially uniform voltage perturbations of high frequency and low magnitude ($-30$ mV). (j) Initiation of sustained ectopic activity in a system with 20\% $I_{KS}$. (k) No initiation of ectopic activity and slow repolarization of quasi steady-state state after high frequency stimulation in a system with 20\% $I_{KS}$.}
\label{fig:hidden_full}
\makeatletter
\let\save@currentlabel\@currentlabel
\edef\@currentlabel{\save@currentlabel(a)}\label{fig:hidden_full:a}
\edef\@currentlabel{\save@currentlabel(b)}\label{fig:hidden_full:b}
\edef\@currentlabel{\save@currentlabel(c)}\label{fig:hidden_full:c}
\edef\@currentlabel{\save@currentlabel(d)}\label{fig:hidden_full:d}
\edef\@currentlabel{\save@currentlabel(e)}\label{fig:hidden_full:e}
\edef\@currentlabel{\save@currentlabel(f)}\label{fig:hidden_full:f}
\edef\@currentlabel{\save@currentlabel(h)}\label{fig:hidden_full:h}
\edef\@currentlabel{\save@currentlabel(g)}\label{fig:hidden_full:g}
\edef\@currentlabel{\save@currentlabel(i)}\label{fig:hidden_full:i}
\edef\@currentlabel{\save@currentlabel(j)}\label{fig:hidden_full:j}
\edef\@currentlabel{\save@currentlabel(k)}\label{fig:hidden_full:k}
\makeatother
\end{figure}

As described earlier, a multiple attractors mechanism explains threshold effects. However it does not clarify  the  frequency selection process. In particular, existence of multiple attractors cannot explain the failure to induce of ectopic activity or to terminate ongoing ectopic activity after high-frequency stimulation. To elucidate this mechanism, let us analyze long trains of high-frequency stimulation from the numerical experiments shown in Figs. \ref{fig:resonance_switch_on} and \ref{fig:termination}. Analysis of long trains is straightforward, since it allows us to eliminate transient effects. Figs. \ref{fig:hidden_full:a} and \ref{fig:hidden_full:b} show signals from the center of the irradiated area during stimulation with 20 and 40 high-frequency  pulses, respectively. As we can see in Fig. \ref{fig:hidden_full:a} , the potential at the center of the pattern reaches the quasi-steady state after several stimulation pulses, whereas Fig.  \ref{fig:hidden_full:b} illustrates  that this quasi-steady state  does not change during 20 additional  pulses. The existence of such a quasi-steady state indicates the presence of multistable single cell dynamics in the irradiated area under the action of a weak modulating stimulation current.  This current can be considered weak for the following reasons. Firstly, it is only mediated via gap junctions (diffusional coupling) and we do not have large gradients in the voltage. Secondly, other currents (activity of ion channels) reached steady state due to the long duration of stimulation. To illustrate the point about single cell level multistability under constant irradiation, let us consider an isolated single cell from the center of the irradiated region in Fig. \ref{fig:resonance_switch_on} with additional total steady-state gap-junctional repolarization current, which is applied to the cell in state ST1. Fig. \ref{fig:hidden_full:c} shows standard excitable dynamics for the case of low frequency stimulation. However, after the train of high frequency pulses, the cell switches to a state of high resting potential [Fig. \ref{fig:hidden_full:d}], thereby manifesting bistable properties. It is worth noting, that such bistable dynamics is fundamentally different from standard bistable dynamics in electrophysiology. The latter usually requires the voltage to cross a certain threshold value to switch to another stable state. In our case, the cell needs to integrate several pulses of high frequency stimuli to transition to another stable state. Since the cell exhibits both excitable and bistable dynamics at the same time, we will refer to such type of behaviour as "hidden" bistability. The name implies that bistable behaviour can be concealed by the excitable type of dynamics for low stimulation frequencies. Such a counterintuitive behaviour can be explained by strong time-dependent conductances of ion channels, which vanish in reaching steady state thereby allowing bistability. Indeed, if we plot steady-state IV curve for the cell as shown in Fig. \ref{fig:hidden_full:e}, the curve is N-shaped and crosses zero current values, confirming that the system is bistable without any form of complex multistability. It is important to emphasize that Fig. \ref{fig:hidden_full:e} represents a steady-state IV-curve instead of standard instantaneous IV-curve. In our case we allowed current to reach steady-state for at least 10 s.  Shorter time periods  would  result in elimination of additional steady states by rectification of the N-shape of the curve. The prominent N-shape of the IV curve is actually more important than the fact of crossing zero current points. The N-shape shows that bistability is possible upon application of some range of values of coupling currents, since any constant bias current just trivially shifts the IV curve up or down. Another important feature is that the maximum voltage value of action potentials is higher than the second steady state. It implies that phase space trajectories "circle around" this second equilibrium point in an excitable regime during low frequency stimulation in Fig. \ref{fig:hidden_full:c}. Thus, these trajectories should be able to "circle around" even at high frequency of stimulation, if the size of such trajectories is big enough. Indeed, upon stimulation with larger voltage perturbations than in Fig. \ref{fig:hidden_full:d} (50 mV vs 20 mV)  the cell does not approach  the second stable state as shown in Fig. \ref{fig:hidden_full:f}, even though the stimulation frequency remained the same.  Such an outcome indicates that the success of transition to the depolarized state is determined by the magnitude of voltage perturbation with respect to the equilibrium point. This result from single dynamics has important consequences for our 2D system. Fig. \ref{fig:hidden_full:h} shows that sustained ectopic activity  can not be stopped by spatially uniform high voltage perturbations (50 mV) of high frequency, while uniform intermediate voltage perturbations (0 mV) of the same frequency stop the activity as shown in Fig. \ref{fig:hidden_full:g}.  Furthermore, shocks with the same frequency and even lower voltage (-30 mV) do not stop ectopic activity as shown in Fig. \ref{fig:hidden_full:i}. Analogous results are achieved for ectopy initiation.   Overall, success (or failure) of ectopy termination (or initiation) depends non-monotonically on the magnitude of voltage perturbation. 
\par These results stress the importance of these "hidden" depolarized steady states in the 2D model of "hidden" bistability and explain the frequency selection properties. After high frequency stimulation by travelling waves,  irradiated cells in the center of the region reach the basin of attraction of this depolarized steady state, which slowly dissipates by the action of the coupling current. The latter fact can be illustrated by simulations in a modified model of our system with an "exaggerated" reduction of the repolarization reserve by lowering the $I_{Ks}$ conductance to 20$\%$ of its original value. After this reduction, which does not significantly alter the basic APD of the system and the system still obeys similar bistable dynamics [Fig. \ref{fig:hidden_full:j}] as an unchanged model, thereby belonging to the same class of systems. In Fig. \ref{fig:hidden_full:k}, after a train of high frequency pulses, the center of the region in the modified model reaches some quasi-steady state and slowly repolarizes without evoking any oscillations, thus illustrating the hypothesis about pattern dissipation.
\par The results in Figs. \ref{fig:hidden_full:h}, \ref{fig:hidden_full:g}, \ref{fig:hidden_full:k} provide just a proxy for rigorous mathematical analysis. However, they establish a qualitative link between the "hidden" bistability  on single cell level and the dynamics of a spatial (multi-cellular) system. The results also illustrate the superiority of termination of ectopic activity by travelling waves compared to by spatially uniform perturbations. In case of travelling waves the system reaches automatically the optimal  distance to the quasi-steady state, while for the stimulating by spatially uniform perturbations the amplitude should be accurately tuned.

\subsection{Minimal reaction-diffusion model}
\label{sec:minimal_RD}
\begin{figure}
\centering
\includegraphics[width=1.0\linewidth]{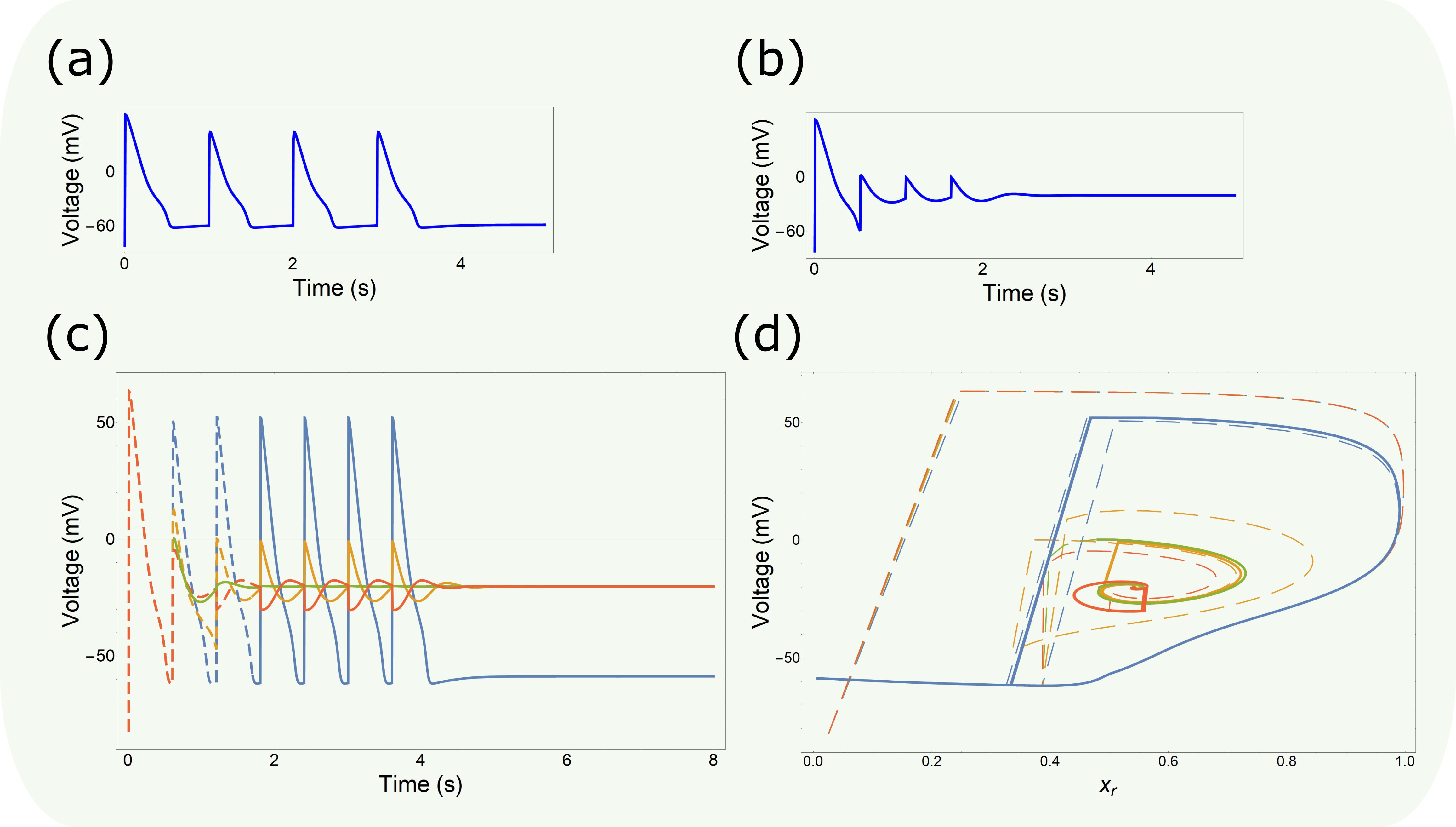}
\caption{"Hidden" bistability in a simplified single cell model. (a) Excitable dynamics for low frequency stimulation. (b) Transition to a depolarized state after a train of high frequency pulses. (c) Evolution of the voltage variable as a result of high frequency voltage perturbations of high (blue trajectories) and low (green, orange and red) magnitudes. The dashed parts of graphs demonstrate the transient non-periodic parts of the trajectories, the solid lines show the periodic and stationary states. (d) Corresponding trajectories in phase plane.}
\label{fig:attractor_reduced}
\makeatletter
\let\save@currentlabel\@currentlabel
\edef\@currentlabel{\save@currentlabel(a)}\label{fig:attractor_reduced:a}
\edef\@currentlabel{\save@currentlabel(b)}\label{fig:attractor_reduced:b}
\edef\@currentlabel{\save@currentlabel(c)}\label{fig:attractor_reduced:c}
\edef\@currentlabel{\save@currentlabel(d)}\label{fig:attractor_reduced:d}
\makeatother
\end{figure}

In the two previous sections, we did not discuss any specific electrophysiological details about the parameters of ionic currents. We rather described generic structures as emergent bistability in a 2D system, its relation to "hidden" bistability on the single cell level under the action of bias current and its sensitivity to the amplitude and frequency of perturbations. We also observed our phenomenon  for a wide range of relevant \textit{in-vitro} parameters.  All these considerations indicate that we deal with a qualitative phenomenon, which may be observable in a wide class of systems besides monolayers of heart muscle cells with regionally activated optogenetic current and which realization is independent of the precise parameters in a concrete system. To illustrate this, we constructed a simple reaction-diffusion model describing our frequency-dependence phenomenon with the minimal possible number of independent variables. To construct the minimal model we systematically eliminated all unnecessary variables and currents while preserving the main "ingredients" of the system such as "hidden" bistability on the single cell level under the action of a bias current  and emergent bistability in 2D. The resulting  system of equations consists of only 3 independent variables (i.e. for voltage, the depolarizing and the repolarizing current) as shown in Supp eq1 and  constant inhomogeneity, which emulates a constant optogenetic current. In this simple system, we were able to reproduce the main important types of dynamics found in the complex system. These types include the resonance phenomenon for transient (state ST3) and sustained ectopy (state ST2) and multistability between ST1, ST2 and ST3 as demonstrated in Figs. Figs. SI3(a), SI3(b), SI3(c), SI3(d). Moreover, similar to what we found in the complex system, sustained ectopic activity could be stopped only by spatially uniform perturbations of intermediate magnitude as shown in  Figs. SI3(e), SI3(f), SI3(h). On the single cell level, inside the inhomogeneity zone and under the action of a constant bias current, the system exhibits the phenomenon of "hidden" bistability [Figs. \ref{fig:attractor_reduced:a}, \ref{fig:attractor_reduced:b}] and responds differently to different amplitudes of stimulation [Figs. \ref{fig:attractor_reduced:c}, \ref{fig:attractor_reduced:d}] in a similar manner to the complex model [Figs. \ref{fig:hidden_full:c}, \ref{fig:hidden_full:d}, \ref{fig:hidden_full:f}. Fig. \ref{fig:attractor_reduced:c} shows time evolution of the voltage variable, Fig. \ref{fig:attractor_reduced:d} demonstrates corresponding trajectories in the phase space of voltage and the state variable of the repolarizing current. The dashed lines in Fig. \ref{fig:attractor_reduced:d}  indicate the transient non-periodic part of the trajectories, the solid lines show the periodic and stationary states. The trajectories are the result of high frequency voltage perturbations of high (blue trajectories) and low (green, orange and red) amplitudes.  Green, red and orange trajectories approach the "hidden" stable depolarized state, which is encircled by blue trajectories of excitable dynamics thus showing coexistence of 2 attractors similar to Figs. \ref{fig:hidden_full:d}, \ref{fig:hidden_full:f}.  Importantly, our model could not be reduced to physically sound 2-variable FitzHugh-Nagumo like models in the form $\frac{d V}{d t}=f(V)+g(n,V), \frac{d n}{d t}=k(n,V)$, where $V$ is voltage, $n$ is some repolarizing variable, $f(V)$   is the nonlinear function for the fast depolarizing current responsible for fast excitation and $g(n,V)$, $k(n,V)$ are functions of the repolarizing current and its kinetics,  respectively. This statement is illustrated by 2D trajectories in the voltage vs repolarizing variable $(V,n)$ plane in Fig. \ref{fig:attractor_reduced:d} for our three-dimensional system.  Thus, if such system is to be realized in a 2-variable model, the fast excitation dynamics should depend both on the voltage and repolarizing variable. Thus  $f(V)$ should change to $f(V,n)$ thereby making the depolarizing current dependent on the inactivation value of the repolarizing current. The latter does not have electrophysiological meaning, since ionic currents primarily act independently of each other's state.

\section{Discussion}
\label{sec:discussion}
In this paper  we exploited the unique feature of optogenetics, i.e. spatiotemporal control of a specific biological function, here excitability,  to reveal and study the emergent mechanisms of frequency dependence of ectopic activity in cardiac tissue. We used  CheRiff, a light-sensitive depolarizing ion channel from the family of channelrhodopsins \cite{wang2015optogenetics} to gradually change the conductance of the background current and study its role in initiating arrhythmia. This allowed us to characterize for the first time the bifurcation pattern of the transition to ectopic activity. We found that this transition involves phases of bistability and multistability and, most excitingly, shows resonant properties. In particular, the possibility to induce ectopic activity depends on its initiation procedure in a resonant manner. In some parametric range ectopic activity can only be initiated if we perturb the system by external travelling waves of a certain frequency range. We also show that such a procedure can  stop sustained ectopic activity if the system is perturbed outside the resonance range.  Observed multistability and resonance are purely emergent properties, which are not present on the single cell level, as directly proved \textit{in-vitro} and \textit{in-silico}. In our \textit{in-vitro} experiments, we observed similar resonant transitions with gradual change of another parameter, the size of the irradiated zone, while keeping the light intensity constant.  This experimental result provides the \textit{in-situ} proof of the emergence and wide parametric range for the observation of the effect. From our \textit{in-silico} findings we can conclude that the emergent rate dependence can be explained by "hidden" bistability on the single cell level in the irradiated zone and the action of repolarizing coupling currents on a monolayer level. Using these two general concepts we were able to construct a generic reaction-diffusion system with a minimal number of independent variables. This minimal system of 3 variables reproduced all dynamical regimes of the complex model, thus displaying the proposed mechanism as generic and not dependent on specific electrophysiological details of the system. The fundamental nature of this mechanism and its concrete \textit{in-vitro} realization under diverse conditions allow us to discuss the phenomenon in a variety of contexts.  Below,  we compare  our phenomenon to different types of arrhythmic events and other generic mechanism of bistability in cardiac tissue, while extending the relevance of our  mechanism to resonant behaviour in neuronal and other biophysical systems. 

\subsection{Resonance and ectopic activity in the context of cardiology}

In connection to the heart, generation of ectopic beats is considered as the main mechanism triggering cardiac arrhythmias. Also, sustained ectopic (focal) activity is  among the two  main mechanisms of maintenance of cardiac arrhythmias \cite{pogwizd1992reentrant}. In most cases, ectopic sources are believed to be a result of triggered activity, which can manifest itself as early afterdepolarizations (EADs) or delayed afterdepolarizations  (DADs) \cite{qu2014nonlinear}. EADs and DADs can directly generate ectopic waves due to  single cell dynamics. Resonant-like frequency-dependence of EADs  can be explained purely on a single cell basis without involvement of collective mechanisms \cite{Tran2009,qu2014nonlinear}.
Initiation of oscillations in our case is  similar to the process of EAD generation, however here it is the result of spatial dynamics of the system. In normal cardiac tissue the optogenetic current is absent but depolarization can potentially be caused by other late activating inwards currents, which are also essential in the induction of EADs or DADs \cite{song2008increase,qu2014nonlinear}. For example, it can be   the  late Na current \cite{belardinelli2015cardiac},  or a  result of reactivation of ICaL \cite{Tran2009}, NCX activity \cite{song2015calcium} or other depolarizing currents. In addition our phenomenon might have relevance to depolarization-induced automaticity or triggered activity caused by background current. Such situation can be observed under condition of oxidative stress, ischemia and chronic heart failure \cite{belardinelli2015cardiac,shryock2013arrhythmogenic,undrovinas1999repolarization}. We propose to probe frequency-induced arrhythmia in these cases. Moreover, our findings might explain the phenomenon of bradycardia-induced triggered activity \cite{brachmann1983bradycardia}. Overall, the collective phenomenon of resonant-like transition to ectopy can be applied to multiple clinical scenarios.

Such sustained oscillations can be explained as a result of the following process. A single uncoupled illuminated cell is in a steady depolarized state due to the additional inward current produced by the  activation of CheRiff.  At the same time, there is also an electrotonic 
coupling repolarizing current due to interaction between depolarized and nondepolarized parts in the coupled system.
If the area of illumination is big, the central cells do not “sense” repolarized distant neighbours (small electrotonic current) resulting in the absence of oscillations in the central region. If the area becomes smaller the effect of electrotonic currents increases, while the optogenetic current remains the same.  As a result   electrotonic currents can repolarize  illuminated  cells and together with the optogenetic current induce  oscillations  of high amplitude in the center. This mechanism can work only if the optogenetic depolarization is not very strong, otherwise the electrotonic current will not be able to repolarize the illuminated cells. On the other hand, if the electrotonic current is too strong it will prevent depolarization by the optogenetic current. As a result, for strong illumination the oscillations will occur closer to the boundary as it will require a stronger electrotonic current. In contrast, for lower illumination intensities, the boundary of oscillation will be shifted toward the center of the illuminated area. As in our experiments, the light intensity was relatively weak and oscillations occurred in the center, i.e. far from the boundary of the illuminated area. However, if illumination is strong, the area of oscillations shifts towards the boundary. Here other factors can become essential. Since one of the factors affecting electrotonic current is the curvature of the boundary between illuminated and not-illuminated regions, the electrotonic current will be the strongest at the corners. This can explain  why we saw firing from the corners  in a previous study in which a strong depolarizing force in the illuminated region \cite{teplenin2018paradoxical}. We observed similar effect of ectopy initiation from the corners in Fig. \ref{fig:size_resonance:b}. However, it is just an auxiliary effect in this paper, used to guarantee monostable single cell behaviour in order to observe bistability and resonant transitions while changing the size of the pattern. 

\subsection{"Hidden" bistability and collective frequency dependence mechanism}
Overall, we were able to achieve resonant conditions under weak and strong optogenetic depolarization of differently-sized areas of illumination.  This wide parametric range is indicative of the fundamental and qualitative  nature of the effect, which was confirmed by our \textit{in-silico} data. The mechanism of resonant transitions was explained by two main processes: 1) accumulation of perturbations at intermediate frequencies; 2) interaction of "hidden" bistable dynamics with coupling currents at high frequencies.  The second process is of extreme importance, since it is responsible for resonant behaviour by cutting off higher frequencies.  "Hidden" bistability of a single cell is a novel phenomenon and deserves separate consideration. Although its manifestations are clear in a monolayer system, "hidden" bistability might be challenging to be observed in single cell patch-clamp experiments. In Fig. \ref{fig:hidden_full:e} the height of dips and hills of the N-shaped steady-state IV curve are around $0.2 pA/pF$. Measurements of these details can be hindered by the statistical and instrumental errors of voltage-clamp recordings.  Voltage bistability can be lost via a shift in current balance due to the non-zero resistance of patch pipette in current clamp mode.  This might explain why such phenomenon thus far has only been reported once in an \textit{in-silico} study dedicated to myotony in muscle fibers  \cite{cannon1993theoretical}.     

The seemingly fragile nature of the phenomenon might be a requirement for coexistence of excitable and bistable regimes. Otherwise, if the hills and dips of N-shaped steady-state IV curve are larger, they would be in the same range as the big currents of fast ion channels thus converting "hidden" bistability to normal bistability. Furthermore, the smallness of the dips and hills might explain the resonant properties. Indeed, small repolarizing coupling currents would be able to switch from quasi-stable depolarized state in Fig. \ref{fig:hidden_full:a} to a repolarized single cell state, corresponding to quiescent state ST2.
The importance of a quasi-stable depolarized state was demonstrated with our \textit{in-silico} data on high frequency voltage perturbations applied to the monolayer system in the experiments focusing on the termination of sustained activity shown in Figs. \ref{fig:hidden_full:h}, \ref{fig:hidden_full:g}, \ref{fig:hidden_full:i}. These data on oscillation termination indicate that our system is fundamentally different from a system with the bistability between limit cycle and resting state described in a classical experiment on single pulse pacemaker annihilation \cite{winfree2001geometry}. Such classical pacemaker annihilation can be achieved by applying sufficiently high frequency voltage perturbations.  These voltage perturbations can be thought of as a purely hypothetical system of interswitching depolarizing and repolarizing currents with the feedback loop control allowing to reach certain voltages in a short period of time. If these voltages lie within the basin of attraction of the resting state, the system will go to the resting state after some number of such voltage perturbations. As discussed above, we do not need to apply perturbations close to resting state ST2 in our system in order to terminate ectopic activity. Also, the appropriate amplitude of perturbation can be "automatically" selected by a high frequency train of travelling waves without any \textit{a priory} knowledge about the state of the system. 
Moreover, stimulation by high frequency waves is the only viable option for arrhythmia  termination, since uniform high voltage shocks cannot terminate sustained activity as shown in Fig. \ref{fig:hidden_full:h}.
As discussed earlier, all key aspects of the aforementioned behaviour can be captured in a simplified 3 variable reaction-diffusion model implicating generality of the effect. However, in the current paper we provide just a proxy for a rigorous mathematical analysis. A full mathematical study  might be challenging even for the simplified equations due to the lack of efficient analytical tools to describe fully developed limit cycle oscillations in reaction-diffusion systems \cite{sherratt2008periodic}. Existing ones are arduous in practice and require elements of advanced algebraic topology, such as the Conley index \cite{kuehn2015multiple}.

\subsection{Relation to resonance in other excitable systems}
Resonance phenomena are well known in   nonlinear biophysical systems. For example, the actomyosin cortex in social amoeba \textit{Dictyostelium discoideum} responds to short pulses of cAMP by damped oscillations in average actin content, thereby allowing fine time-selective responses to either internal (pseudopodia initiation) or external(cAMP release) perturbations  \cite{westendorf2013actin}. In  synthetic gene regulatory networks, subthreshold resonance might be used to discriminate environmental stimuli (e.g. growth factor release, heat shock) \cite{guantes2006dynamical}. Also  neurons can  differentiate  external stimuli depending on their frequency. 
It is interesting to compare the resonance properties of our system with those actively studied in  neuroscience.  \par
In neuroscience,  the resonance phenomenon is usually associated with type II excitability governed by so-called membrane resonance on the single cell level or, mathematically by subthreshold oscillations initiated due to proximity of the resting state to a Hopf bifurcation \cite{hutcheon2000resonance,IZHIKEVICH2000}. Such resonant neurons, which are referred to as resonate-and-fire neurons in mathematical neuroscience \cite{IZHIKEVICH2000}, are of paramount importance for some  neurocomputational properties \cite{roach2018resonance}. An important property of these neurons is that the resting state  lies inside the limit cycle.  However, in our case the resting state is stable and does not lie inside the limit cycle. In Fig. SI2 we show six 2D projections of trajectories (limit cycles) corresponding to sustained ectopic activity (state ST1) and to a steady  state ST2, which lies outside the limit cycle. This result was corroborated by time-delayed embedding analysis \cite{takens1981detecting} of our experimental data in Fig. \ref{fig:explanation_threshold_exp}.  Another property of our system is that it requires several external waves to initiate sustained ectopic activity. This is similar to properties of  neurons of excitability type I  (so-called integrate-and-fire neurons) \cite{IZHIKEVICH2000}, which is the  model of first choice for single neurons in computational neuroscience. These neurons do require accumulation of perturbations to fire. However, compared to our systems they do not show frequency preference. Thus our system shows an intermediate type between the resonate-and-fire and integrate-and-fire neurons. These types of spiking neurons are usually associated with generic Bogdanov-Takens bifurcation \cite{IZHIKEVICH2000}, however, another type of bifurcation, saddle-node-loop (SNL) bifurcation, is also capable to generate spiking neurons. The latter bifurcation allows a bistability regime, in which the stable state lies outside the limit cycle as in our case in Fig. SI2.  \par 
Spike generating SNL bifurcation received renewed attention \cite{hesse2017qualitative,hesse2019correctly,schleimer2021firing} and investigation of this mechanism on a network level  is a new line of research \cite{website}.
Interestingly, SNL bifurcation was also identified in the model of transcriptional regulation of the cell \cite{p53_mdm2} and recently in a novel design of combined genetic switches \cite{PEREZCARRASCO2018521}. Our result might represent the emergent analog of SNL bifurcation with resonant interstate transitions, which also might be probed on the level of neuronal network or general nonlinear networks. Results from the literature on complex networks suggest plausibility of such scenario. SNL bifurcation has already been identified in Kuramoto type of models in the networks of coupled limit cycle oscillators both in broad \cite{PhysRevE.79.026204,childs_snl} and in neuroscience context \cite{PhysRevResearch.3.023224}. These type of networks can exhibit resonant phenomenon in addition to SNL bifurcation for a couple of reasons. First, in these studies SNL bifurcation was identified using dimensionality reduction techniques such as Ott-Antonsen ansatz in the limit of infinite number of oscillators, while ignoring the whole wealth of transient and finite-size phenomenon.  Second, the Kuramoto model is an idealized model and does not contain many physical details of the system. It describes only a weak forcing regime on an idealized limit cycle oscillator, in which oscillations can not be destroyed but only phase shifted. In our case, we provide a more realistic biophysically motivated model, which allows a strong forcing regime and destruction of oscillations due to the phenomenon of "hidden" bistability. Both arguments give us substantial reasons to believe that aforementioned idealized Kuramoto models can possibly possess resonant properties if augmented/modified with sufficient details from corresponding full physical/biophysical models.

\par
On a more speculative note, our level of biophysical description can also help to provide an alternative to a recently proposed mathematical framework explaining success of high frequency stimulation for termination of neuronal oscillations during Parkinson disease \cite{PhysRevX.10.011073}. There, authors use network models with elementary units governed by two variable dynamics to describe termination mechanisms. Here, we show that a model of elementary unit with no less than three variable is necessary to capture the termination mechanism, since the phenomenon of "hidden" bistability requires at least three independent variables.
\par 
In our system, resonance phenomena are evoked by external propagating waves. Recently,  multicompartment cable computational models of the neuron  were introduced in neuroscience \cite{schwemmer2012bistability}.  As such, these models are spatial and can also show resonant properties similar to those described in our paper, where resonant phenomena are the collective effect of spatial dynamics. Such possibilities have not yet been discussed  in the context of resonance phenomena. We hope that  our findings  provide an impetus to search for resonance in multicompartment cable computational models of the neuron, since bistablity between quiescence and spiking  has already been found in neurons \textit{in-silico} \cite{schwemmer2012bistability} as a purely emergent phenomenon, and also \textit{in-vitro} \cite{le2006bistable}.  Our findings may also trigger studies into finding similar phenomena in neuroscience by local optogenetic activation of depolarizing ion channels, both \textit{in-vivo} \cite{Lu2015a} and \textit{in-silico} in nonlocal models  \cite{Selvaraj2015,Heitmann2017}. 
\par
In conclusion, we have demonstrated, using \textit{in-vitro} and \textit{in-silico} models, a fundamentally new emergent mechanism for resonance in non-homogeneous excitable systems. This mechanism was shown to be responsible for triggering and terminating certain types of ectopic arrhythmias. Given the generality of the description, this new mechanism of resonance could be used to study or control collective states in other excitable systems.
\begin{acknowledgments}
\noindent \textbf{Funding:} This work was supported by the Netherlands Organization for Scientific Research (NWO Vidi grant 917143 to D.A.P.)  and European Research Council (ERC starting grant 716509 to D.A.P.). A.V.P. was funded  by the Ministry of Science and Higher Education of the Russian Federation within the framework of state support for the creation and development of World-Class Research Centers "Digital biodesign and personalized healthcare" 075-15-2020-926.\\
\noindent \textbf{Author Contributions:} A.S.T. initiated and designed research, performed \textit{in-vitro} experiments and computations, analyzed data, conceived theoretical explanations. D.A.P. curated the data collection, A.V.P. guided the theoretical analysis. R.M. and N.N.K. provided CUDA code for Majumder-Korhonen model. A.A.F.V. designed genetic engineering tools. A.A.F.V. and D.A.P. provided reagents and funding sources. A.S.T., N. N. K., A.V.P., A.A.F.V. and D.A.P. wrote the manuscript.\\
\end{acknowledgments}



%

\end{document}